\documentclass[useAMS,usenatbib]{mn2e}

\usepackage{graphicx,amssymb,amstext} 

\def\aj{AJ}			
\def\araa{ARA\&A}		
\def\apj{ApJ}		
\def\apjl{ApJ}		
\def\apjs{ApJS}				
		
\def\aap{A\&A}

\def\mnras{MNRAS}
\def\nat{Nat}

\def\pasp{PASP}

\def\physrep{Phys. Rep.}

\begin{document}

\title[Connecting stellar mass and star-formation rate to dark matter halo mass]{Connecting stellar mass and star-formation rate to dark matter halo mass out to $z\sim2$}
\author[L.~Wang et al.]
{\parbox{\textwidth}{\raggedright L.~Wang,$^{1, 2}$\thanks{E-mail: \texttt{lingyu.wang@sussex.ac.uk}}
D.~Farrah,$^{1, 3}$
S.J.~Oliver,$^{1}$
A.~Amblard,$^{4}$
M.~B{\'e}thermin,$^{5,6}$
J.~Bock,$^{7,8}$
A.~Conley,$^{9}$
A.~Cooray,$^{7,10}$
M.~Halpern,$^{11}$
S.~Heinis,$^{12}$
E.~Ibar,$^{13}$
O.~Ilbert,$^{12}$
R.J.~Ivison,$^{13,14}$
G.~Marsden,$^{11}$
I.G.~Roseboom,$^{1,14}$
M.~Rowan-Robinson,$^{15}$
B.~Schulz,$^{7,16}$
A.J.~Smith,$^{1}$
M.~Viero$^{7}$ and
M.~Zemcov$^{7,8}$}\vspace{0.4cm}\\
\parbox{\textwidth}{\raggedright $^{1}$Astronomy Centre, Dept. of Physics \& Astronomy, University of Sussex, Brighton BN1 9QH, UK\\
$^{2}$Institute for Computational Cosmology, Department of Physics, Durham University, Durham, DH1 3LE, UK\\
$^{3}$Department of Physics, Virginia Tech, Blacksburg, VA 24061, USA\\
$^{4}$NASA, Ames Research Center, Moffett Field, CA 94035, USA\\
$^{5}$Laboratoire AIM-Paris-Saclay, CEA/DSM/Irfu - CNRS - Universit\'e Paris Diderot, CE-Saclay, pt courrier 131, F-91191 Gif-sur-Yvette, France\\
$^{6}$Institut d'Astrophysique Spatiale (IAS), b\^atiment 121, Universit\'e Paris-Sud 11 and CNRS (UMR 8617), 91405 Orsay, France\\
$^{7}$California Institute of Technology, 1200 E. California Blvd., Pasadena, CA 91125, USA\\
$^{8}$Jet Propulsion Laboratory, 4800 Oak Grove Drive, Pasadena, CA 91109, USA\\
$^{9}$Center for Astrophysics and Space Astronomy 389-UCB, University of Colorado, Boulder, CO 80309, USA\\
$^{10}$Dept. of Physics \& Astronomy, University of California, Irvine, CA 92697, USA\\
$^{11}$Department of Physics \& Astronomy, University of British Columbia, 6224 Agricultural Road, Vancouver, BC V6T~1Z1, Canada\\
$^{12}$Laboratoire d'Astrophysique de Marseille, OAMP, Universit\'e Aix-marseille, CNRS, 38 rue Fr\'ed\'eric Joliot-Curie, 13388 Marseille cedex 13, France\\
$^{13}$UK Astronomy Technology Centre, Royal Observatory, Blackford Hill, Edinburgh EH9 3HJ, UK\\
$^{14}$Institute for Astronomy, University of Edinburgh, Royal Observatory, Blackford Hill, Edinburgh EH9 3HJ, UK\\
$^{15}$Astrophysics Group, Imperial College London, Blackett Laboratory, Prince Consort Road, London SW7 2AZ, UK\\
$^{16}$Infrared Processing and Analysis Center, MS 100-22, California Institute of Technology, JPL, Pasadena, CA 91125, USA}}

\date{Accepted . Received ; in original form }

\maketitle

\begin{abstract}

We have constructed an extended halo model (EHM) which relates the total stellar mass and star-formation rate (SFR) to halo mass ($M_{\rm h}$). An empirical relation between the distribution functions of total stellar mass of galaxies and host halo mass, tuned to match the spatial density of galaxies over $0<z<2$ and the clustering properties at $z\sim0$, is extended to include two different scenarios describing the variation of SFR on $M_{\rm h}$. We also present new measurements of the redshift evolution of the average SFR for star-forming galaxies of different stellar mass up to $z=2$, using data from the {\it Herschel} Multi-tiered Extragalactic Survey (HerMES) for infrared-bright galaxies.
%the SFR - $M_{\rm h}$ relation has a bump between a few times $10^{11}M_{\odot}$ and a few times $10^{12} M_{\odot}$;

% An empirical relation between the distribution functions of total stellar mass of galaxies and host halo mass, tuned to match data over $0<z<2$, is extended to include two different scenarios describing the variation of SFR on $M_{\rm h}$.  Key datasets used to build the EHM include the Sloan Digital Sky Survey (SDSS), Cosmic Evolution Survey (COSMOS), Multiwavelength Survey by Yale-Chile (MUSYC), All-wavelength Extended Groth strop International Survey (AEGIS) and {\it Herschel} Multi-tiered Extragalactic Survey (HerMES), which is crucial for deriving accurate SFRs for dusty star-forming galaxies at high redshift. 

Combining the EHM with the halo accretion histories from numerical simulations, we trace the stellar mass growth and star-formation history in halos spanning a range of masses. We find that: (1) The intensity of the star-forming activity in halos in the probed mass range has steadily decreased from $z\sim2$ to $0$; (2) At a given epoch, halos in the mass range between a few times $10^{11}M_{\odot}$ and a few times $10^{12} M_{\odot}$ are the most efficient at hosting star formation;  (3) The peak of SFR density shifts to lower mass halos over time; (4) Galaxies that are forming stars most actively at $z\sim2$ evolve into quiescent galaxies in today's group environments, strongly supporting previous claims that the most powerful starbursts at $z\sim2$ are progenitors of today's elliptical galaxies.

\end{abstract}


\section{INTRODUCTION}

In the past 20 years or so, impressive progress has been made in characterising the evolution of galaxy physical properties over a large fraction of cosmic time. A consistent picture, at least crudely, has emerged in which the global stellar mass density decreases by a factor of 2 or so from $z\sim0$ to 2 and the comoving cosmic star-formation rate (SFR) density increases by more than a factor of 10 over the past 8 Gyr, peaks around $z\sim2$ to 3 and then declines almost linearly with time to higher redshift (e.g., Lilly et al. 1996; Madau et al. 1998; Rudnick et al. 2003;  Dickinson et al. 2003; Schiminovich et al. 2005; Hopkins \& Beacom 2006; Arnouts et al. 2007; Pascale et al. 2009; Bouwens  et al. 2010). The key question that dominates both observational and theoretical efforts today is what physical processes play the dominant role in driving the evolution of the cosmic star-formation activity. Processes such as a decline in the major-merger rate, reduced gas accretion in halos, feedback from central massive black holes and supernova, and  environmental effects (ram pressure stripping of gas, strangulation of the extended halo, etc.) can all impact on the star-formation activity (e.g., Kere{\v s} et al. 2005, 2009; Bell et al. 2005; Bower et al. 2006; Croton et al. 2006; Somerville et al. 2008; Lotz et al. 2008).

Since galaxies form in dark matter halos and their evolution is influenced by the accretion and successive merging of halos (White \& Rees 1978; Fall \& Efstathiou 1980; Blumenthal et al. 1984), it is reasonable to assume that the physical properties of galaxies should correlate to those of the host halos (such as the mass of the halo). Observationally, finding the mass of the host halo can be achieved in a number of ways, e.g., weak gravitational lensing (McKay et al. 2001; Hoekstra et al. 2004; Sheldon et al. 2004; Mandelbaum et al. 2006; Sheldon et al. 2009), dynamical measurement of satellite galaxies (McKay et al. 2002; van den Bosch et al. 2004; Conroy et al. 2007;  More et al. 2011) and X-ray studies (Lin et al. 2003; Lin \& Mohr 2004; Vikhlinin et al. 2006). These techniques are at present expensive in terms of observing time and limited to low $z$ and small dynamical range in halo mass. 

Alternatively, the halo model provides a simple but powerful way to statistically link galaxies with halos. In its simplest form, the halo occupation distribution (HOD), which gives the probability of finding $N$ galaxies (with some specified properties) in a halo of mass $M_{\rm h}$ is used to interpret galaxy clustering (e.g., Peacock \& Smith 2000; Seljak 2000; Scoccimarro et al .2001; Berlind \& Weinberg 2002; Zehavi et al. 2004). Modifications of the HOD include the conditional luminosity function (CLF) which encodes the number of galaxies as a function of luminosity in a given halo (Yang et al. 2003; van den Bosch et al. 2003; Vale \& Ostriker 2004) and the conditional stellar mass function (CSMF) which encodes the number of galaxies as a function of stellar mass in a given halo (Yang et al. 2009; Moster et al. 2010; Behroozi et al. 2010).

In this paper, we build an extended halo model (EHM) to connect stellar mass, $m_*$, and SFR, $\psi$, with the host halo mass, $M_{\rm h}$. The EHM is a hybrid model composed of a parametrised $m_*$ - $M_{\rm h}$ relation and a non-parametric $m_*$ - $\psi$ relation. The first part of the EHM is to use a parametrised relation between the distribution of stellar mass and halo mass, i.e. the CSMF, to describe the statistical relation between $m_*$ and $M_{\rm h}$. The parameters in the CSMF at $z\sim0$ are tuned by the spatial density and clustering of galaxies while their evolution in the redshift range $0<z<2$ is constrained by galaxy SMFs only. The second part of the EHM is to extend the CSMF to the joint distribution in $m_*$ and $\psi$ as a function of $M_{\rm h}$, using two different scenarios for the role of $M_{\rm h}$ in determining the distribution of $\psi$ at fixed $m_*$. This second part is non-parametric as we use the observed conditional SFR distributions at fixed $m_*$ as direct inputs. The key to building the EHM is a large sample of galaxies with reliable $m_*$ and $\psi$ estimates. The Herschel Multi-tiered Extragalactic Survey (HerMES; Oliver et al. 2011) covering most of the well-studied extragalactic fields with ancillary data from the X-ray to radio is the perfect place to start such a project.

The layout of the paper is as follows. In Section 2, first we briefly describe the published measurements used to constrain the EHM. Then, we describe the data-sets used to derive the stellar masses and SFRs of high-redshift galaxies in HerMES fields. In Section 3, we present the CSMF in both the local Universe and at high redshift. The evolution of the stellar content as a function of $M_{\rm h}$ is derived using the CSMF as a function of redshift and the halo accretion history from N-body simulations. In Section 4, we extend the CSMF to the 2-D distribution of galaxies in the $(\psi, m_*)$ plane as a function of $M_{\rm h}$. The evolution of the star-formation activity as a function of $M_{\rm h}$ is derived using the EHM as a function of redshift and the halo accretion history. Finally, conclusions and discussions are presented in Section 5. Unless otherwise stated, we assume $\Omega_M=0.3$, $\Omega_{\Lambda}=0.7$, $\sigma_8=0.8$ and $h=0.7$. All magnitudes are in the AB system.

\section{Data-sets}

To constrain the $m_*$ - $M_{\rm h}$ relation at $z\sim0$, we use the published stellar mass function (SMF) (Guo et al. 2010) and correlation functions of the SDSS galaxies (Li et al. 2006). To constrain the redshift evolution of the $m_*$ - $M_{\rm h}$ relation, we use the published SMFs in P{\'e}rez-Gonz{\'a}lez et al. (2008) based on a combined sample of 3.6 and 4.5 $\mu$m selected sources in the HDF-N, the CDF-S and the Lockman Hole. The clustering properties of high-redshift galaxies are not used to constrain the evolution parameters in the $m_*$ - $M_{\rm h}$ relation due to issues explained in Section 3.2.
%various redshift bins in the Cosmological Evolution Survey (COSMOS; Scoville et al. 2007) (Ilbert et al. 2010).

To extend the CSMF to the joint distribution in $m_*$ and $\psi$ as a function of $M_{\rm h}$, we use the conditional probability distribution function (PDF) of SFR of the entire population as a function of $m_*$. The conditional PDF of SFR of galaxies in the local Universe is taken from Salim et al. (2007). To derive the conditional SFR distribution as a function of $m_*$ in the distant Universe, we use galaxies observed in three well-studied extragalactic fields,  the Extended Chandra Deep Field-South (ECDFS) field, the COSMOS field and the Extended Groth Strip (EGS). We will describe in detail the data-sets used in each field below.

\begin{figure*}\centering
\includegraphics[height=2.in,width=2.2in]{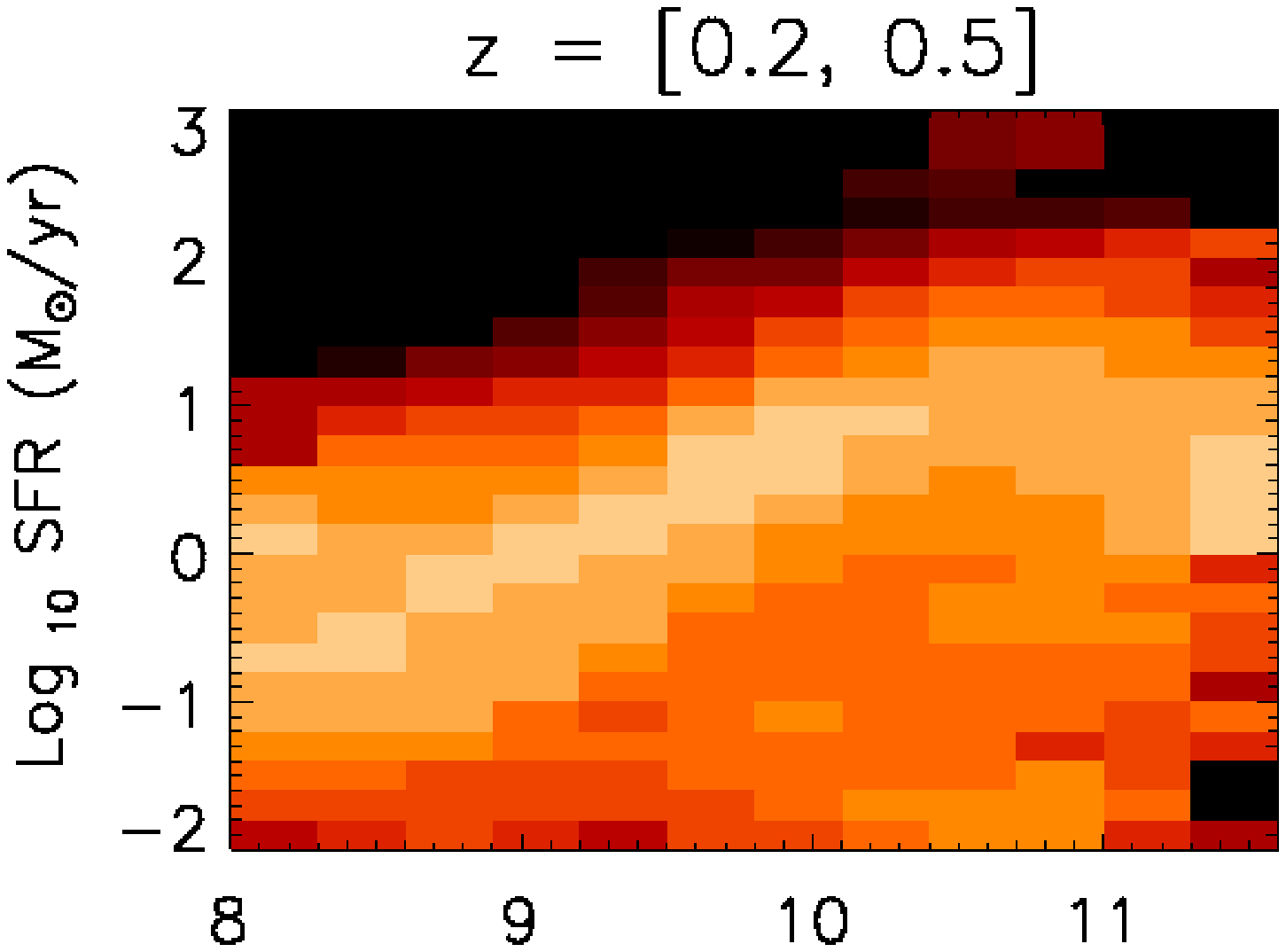}
\includegraphics[height=2.in,width=2.in]{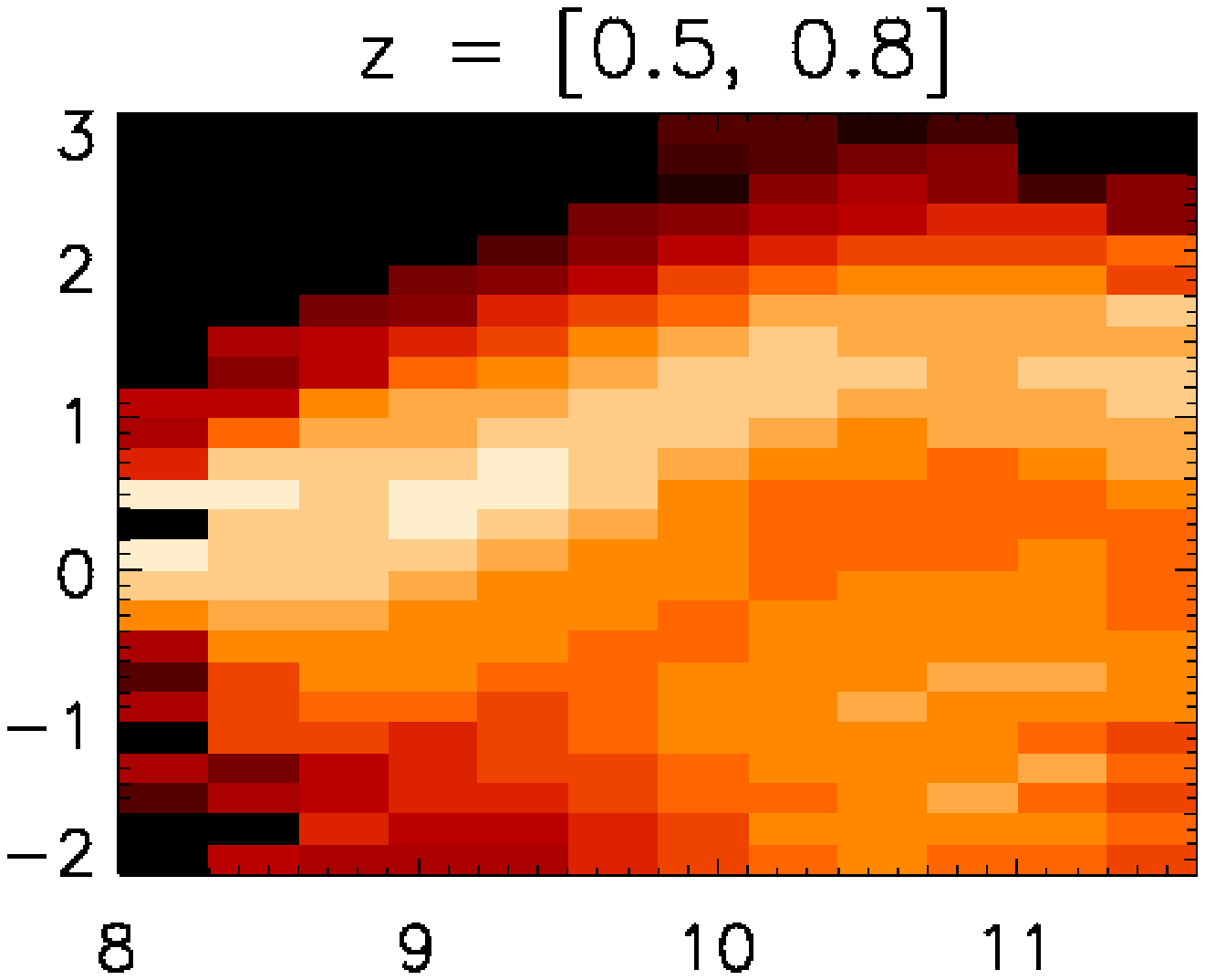}
\includegraphics[height=2.in,width=2.in]{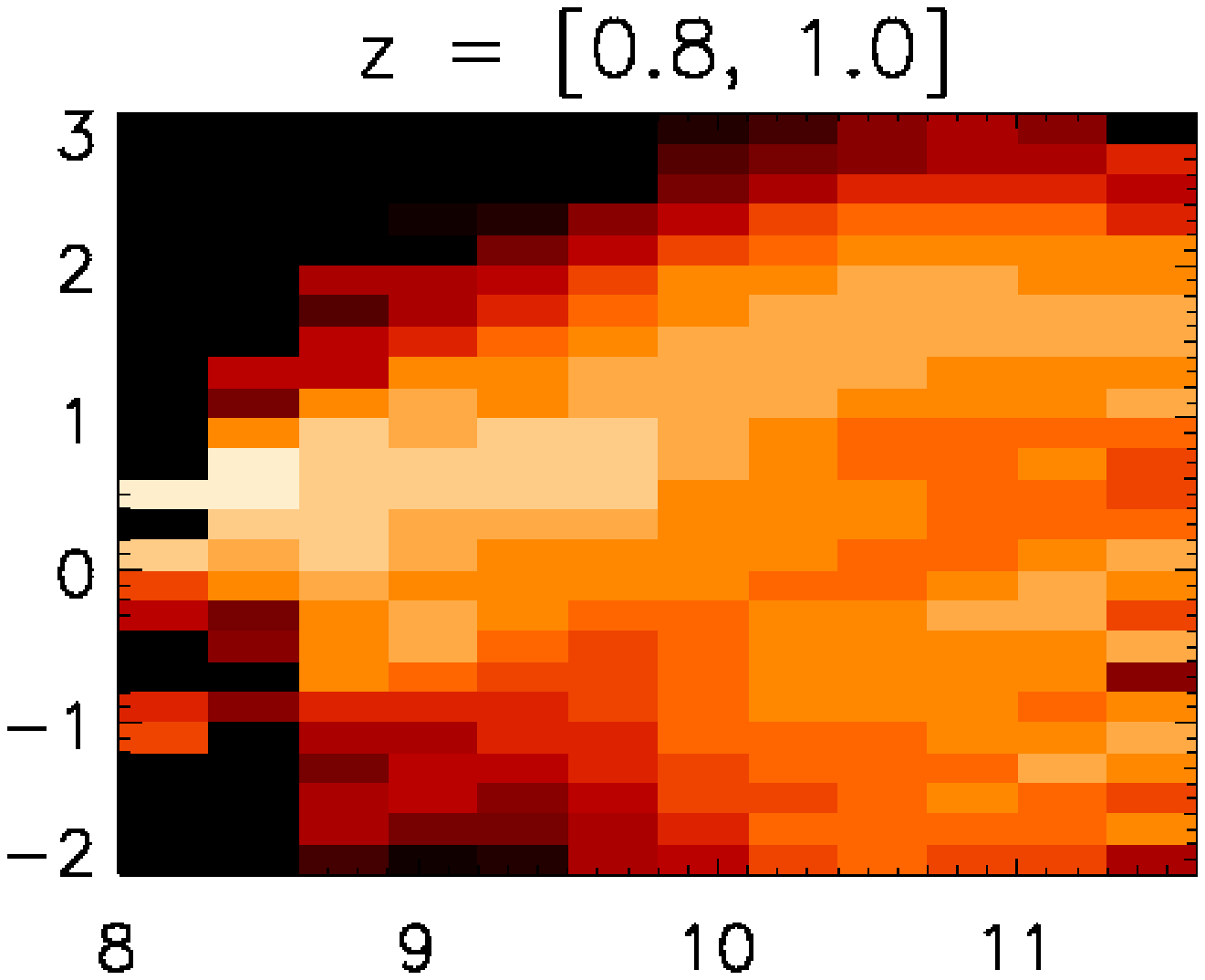}
\includegraphics[height=2.2in,width=2.2in]{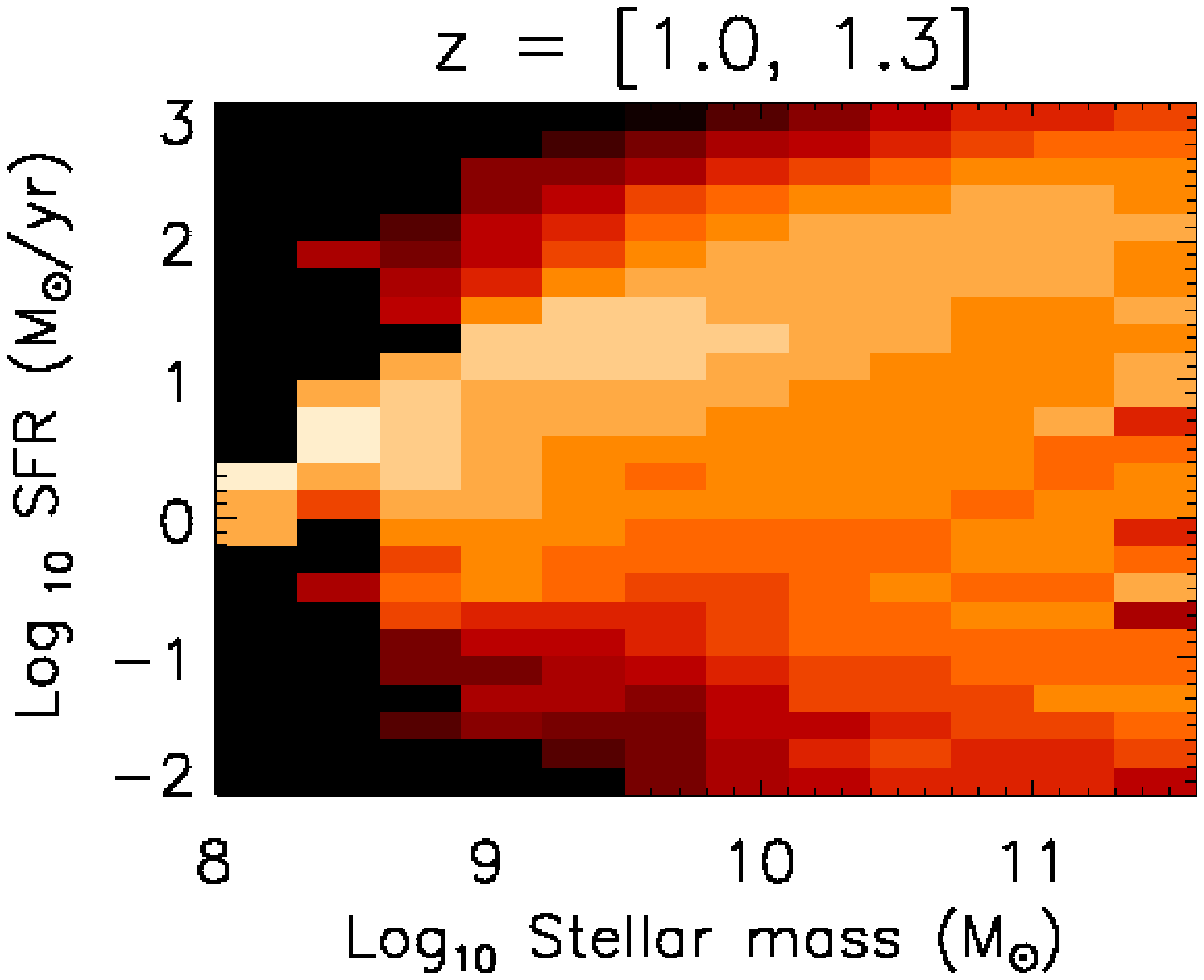}
\includegraphics[height=2.2in,width=2.in]{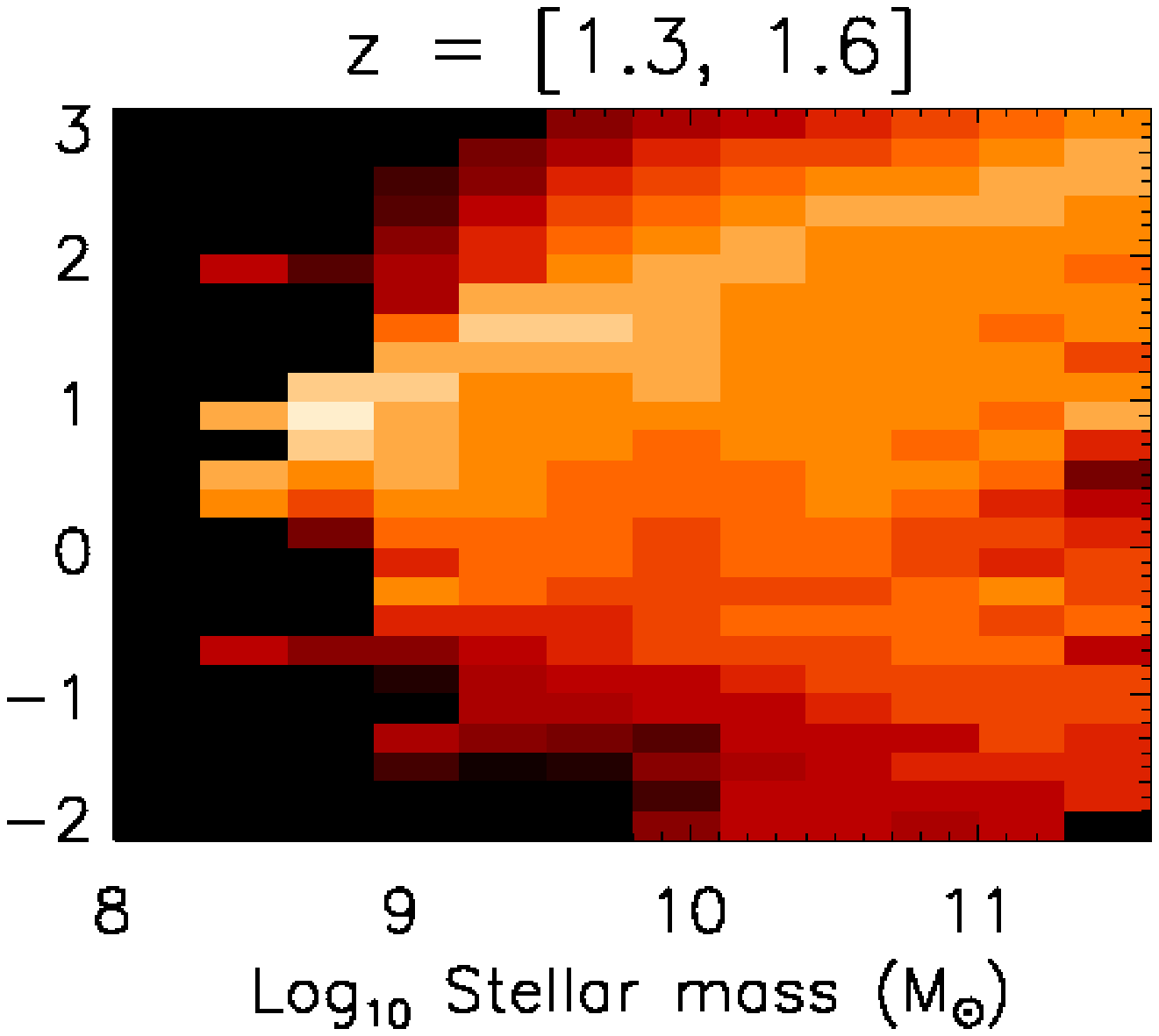}
\includegraphics[height=2.2in,width=2.in]{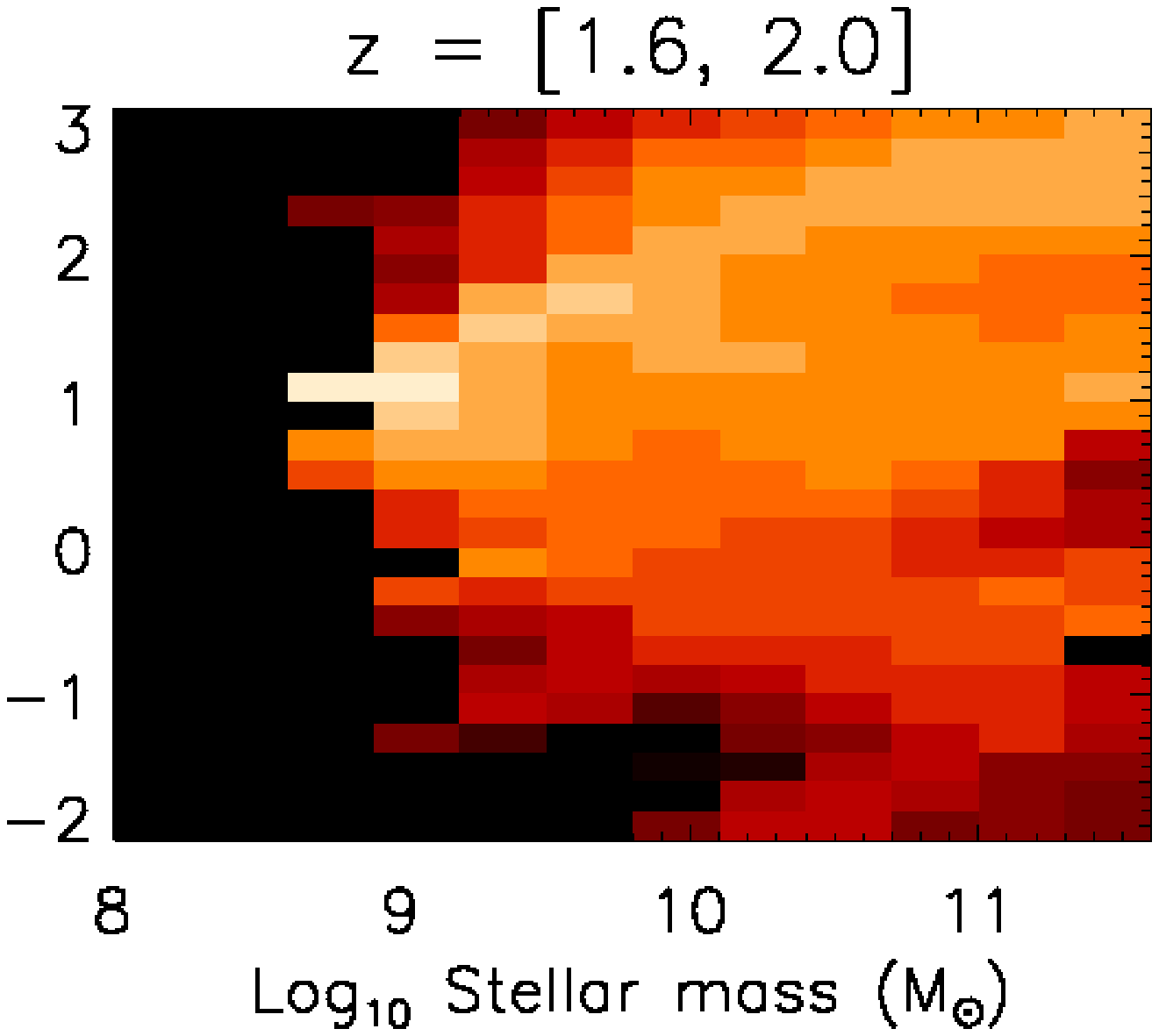}
\caption{The conditional PDF of SFR as a function of stellar mass, in six redshift bins (from left to right and top to bottom, $z=[0.2, 0.5], [0.5, 0.8], [0.8, 1.0], [1.0, 1.3], [1.3, 1.6]$ and $[1.6, 2.0]$), averaged over COSMOS, ECDFS and EGS. The star-forming sequence can be clearly seen in each panel and it evolves upwards roughly independently of stellar mass.}
\label{fig:duty_cycle_conditional_highz}
\end{figure*}

\begin{figure*}\centering
\includegraphics[height=2.3in,width=3.3in]{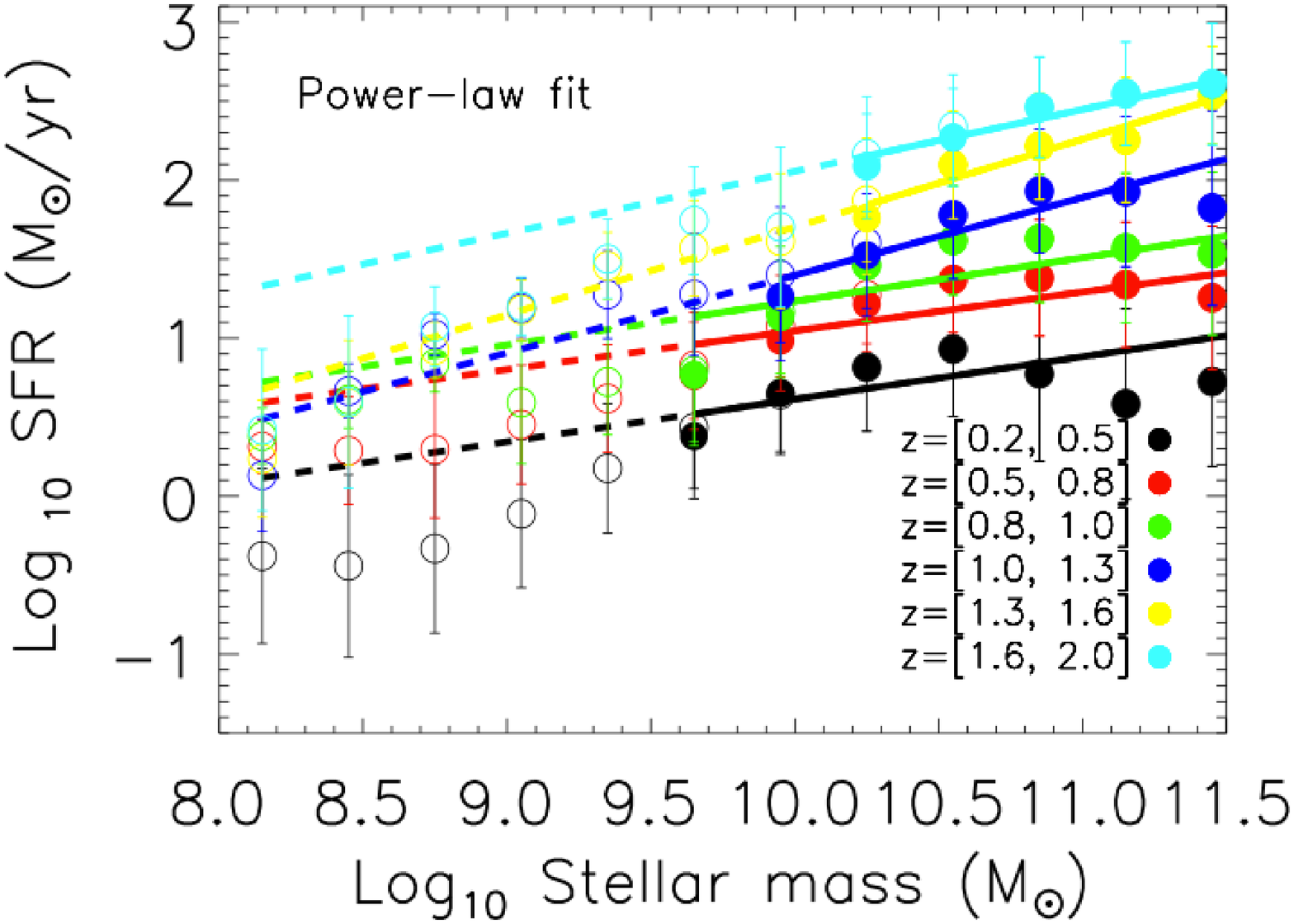}
\includegraphics[height=2.3in,width=3.3in]{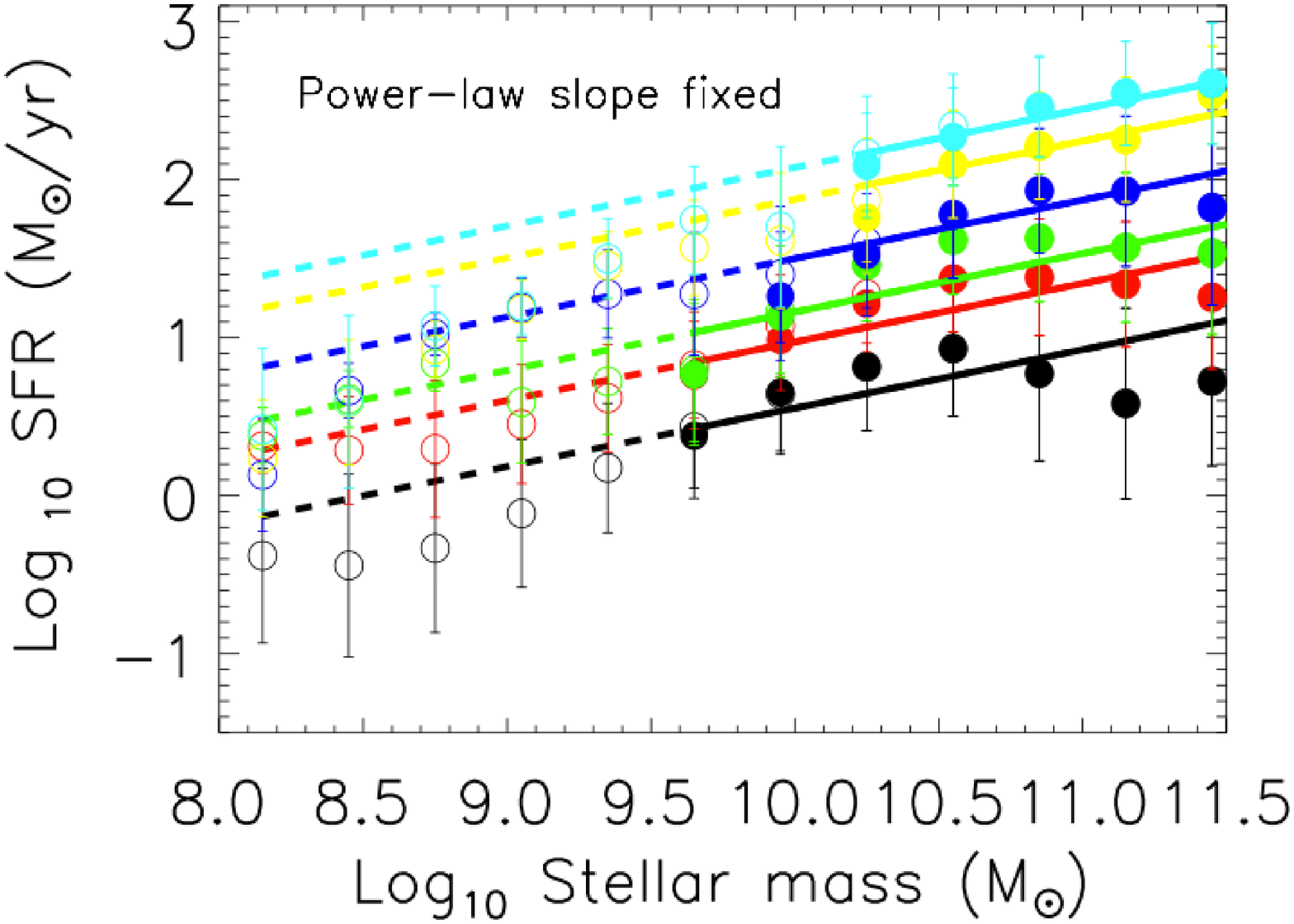}
\caption{The redshift evolution of the average SFR as a function of stellar mass for star-forming galaxies and the best-fit power-law in each redshift slice. The open (filled) circles represent values which are derived from samples below (above) the completeness limit in ECDFS, COSMOS and EGS (see Table 1).  Errors include the field-to-field variations and photometric redshift uncertainty. In the left panel,  both parameters in the power-law are allowed to vary. In the right panel, the power-law slope is fixed to be 0.37 which is the average value over different redshift slices. Note that the power-law fitting is only applied to the filled circles. The best-fit parameters in both panels are listed in Table 2.}
\label{fig:duty_cycle_conditional_highz_fit}
\end{figure*}

\subsection{COSMOS}

The COSMOS photometric redshift catalogue derived from broad and medium bands (GALEX FUV and NUV, optical to infrared data $u^*B_JV_Jg^+r^+i^+i^*z^+JK_sK$, 14 medium and narrow bands from Subaru and 4 IRAC channels) is described in Ilbert et al. (2009). We use an updated version (v1.8 dated from the 13th of July 2010) of Ilbert et al. (2009). The quality of the photometric redshift is very high with $1\sigma$ in  $(1+z)\sim0.007$ at $i_{\rm AB}^+<22.5$.
At $i_{\rm AB}^+<24$. and $z<1.25$, $1\sigma$ in  $(1+z)\sim0.012$. The deep NIR and IRAC coverage enables the photo-$z$ to be extended to $z\sim2$, with $1\sigma$ in $(1+z)\sim0.06$ at $i_{\rm AB}^+ \sim 24$.  Following Ilbert et al. (2010), we construct a mass selected sample as generated from the 3.6 $\mu$m catalogue of the S-COSMOS survey (Sanders et al. 2007). We cross-match the 3.6 $\mu$m and the latest photo-$z$ catalogue by taking the nearest match within 1". The probability of incorrect identification is $<1\%$ (Ilbert et al. 2010).  We then select sources with $f_{3.6} \ge 5 \mu$Jy (the 90\% complete limit), around 2.8$\%$ of which are not matched to an optical counterpart. Using the public photo-$z$ catalogue from the NEWFIRM Medium-band Survey (Whitaker et al. 2011) covering a small area of the COSMOS field but is deeper than Ilbert et al. (2009), we estimate that only $1\%$ ($2.5\%$) of the sources with $f3.6 \ge 5 \mu$Jy lie at $z<1.6$ ($z<2.0$) or do not have a photo-$z$ estimate. As we are only concerned with the relation between $m_*$, $\psi$ and $M_{\rm h}$ at $z<2$, we will ignore this $1\%$ of 3.6 $\mu$m sources in our analysis.

\subsection{ECDFS}

We use the Multiwavelength Survey by Yale-Chile (MUSYC) Subaru v1.0 Catalog (Cardamone et al. 2010) containing over 84400 sources.  The catalog includes photometry in 32 MUSYC images of the ECDF-S region, including optical to infrared data (UU38BVRIzJHK), 18 medium bands from Subaru and 4 IRAC channels as part of the SIMPLE survey (Damen et al. 2010), for all sources detected in the combined BVR image. Photometric redshifts are determined using the EASY code (Brammer et al. 2008). The quality of the photometric redshifts is very high, with $1\sigma=0.007$ in $(1+z)$ in the $z=[0.1, 1.2]$, similar to that of the COSMOS field. At $z=[1.2, 3.7]$,  the photometric redshift accuracy gets worse with $1\sigma=0.02$ in $(1+z)$. We select $3.6$ $\mu$m sources above the completeness limit which is $1\mu$Jy.

%sm_limit_cosmos = [9.8, 10.15, 10.05, 10.2, 10.4, 10.65];;the stellar mass limits in each field
%sm_limit_ecdfs  = [9.5,   9.5, 9.5,   9.65, 9.85, 10.05] 
%sm_limit_egs    = [9.92, 10., 10.05, 10.11, 10.29, 10.55]

\begin{table}
\caption{Stellar-mass-selected samples in COSMOS, EGS and ECDFS. The columns are redshift range and stellar mass $m_*$ limit in each field above which our samples are regarded as representative.}\label{table:mass_samples}
\begin{tabular}[pos]{llll}
\hline
\hline
        & COSMOS & ECDFS & EGS \\
  $z$  range         & $\log (m_{\star} [M_{\odot}])$& $\log (m_{\star} [M_{\odot}])$ & $\log (m_{\star} [M_{\odot}])$ \\
\hline
\hline
$z=[0.2, 0.5]$      &  9.8   &  9.5 & 9.9\\
 $z=[0.5, 0.8]$     & 10.1  & 9.5 &10.0\\
 $z=[0.8, 1.0]$     &  10.1  & 9.5& 10.1\\  
$z=[1.0, 1.3]$      &  10.2  & 9.7 & 10.1\\
$z=[1.3, 1.6]$      &   10.4 & 9.9  & 10.3\\
 $z=[1.6, 2.0]$     &   10.7 & 10.1 & 10.6\\ 
\hline
\end{tabular}
\end{table}

\subsection{EGS}

We use an 3.6 + 4.5 $\mu$m selected catalogue in the Extended Groth Strip (EGS) containing 28-band photometry from the ultraviolet to the far-infrared (GALEX FUV and NUV, CFHTLS $u^*g'r'i'z'$, MMT $u'giz$, CFHT12k BRI, ACS $V_{606} i_{814}$, Subaru R, NICMOS $J_{110}H_{160}$, MOIRCS $K_s$, CAHA $J K_s$, WIRC $JK$ and 4 IRAC channels) (Barro et al. 2011a, 2011b). The typical photometric redshift accuracy is $1\sigma=0.034$ in $(1+z)$, with a catastrophic outlier fraction of 2\%. We apply the $90\%$ completeness limit at 3.6 $\mu$m by selecting sources with $f3.6\leq2.3\mu$Jy over areas with homogeneous depth $52.025^{\circ} \leq \delta \leq 53.525^{\circ}$. We also mask out regions in the wings of bright stars.

\subsection{Deriving stellar mass and SFR from HerMES and ancillary data}

\begin{figure*}\centering
\includegraphics[height=2.6in,width=3.35in]{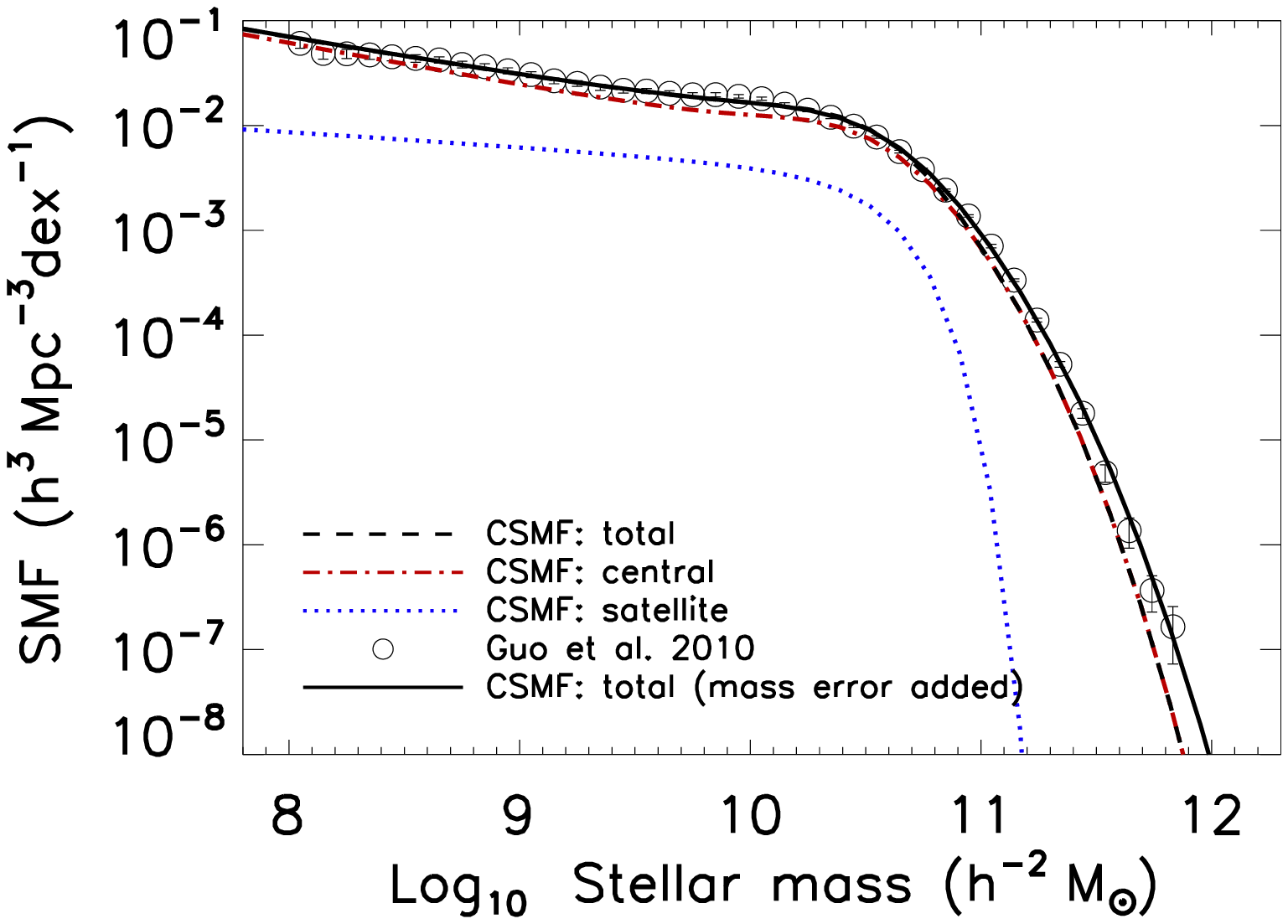}
\includegraphics[height=2.7in,width=3.45in]{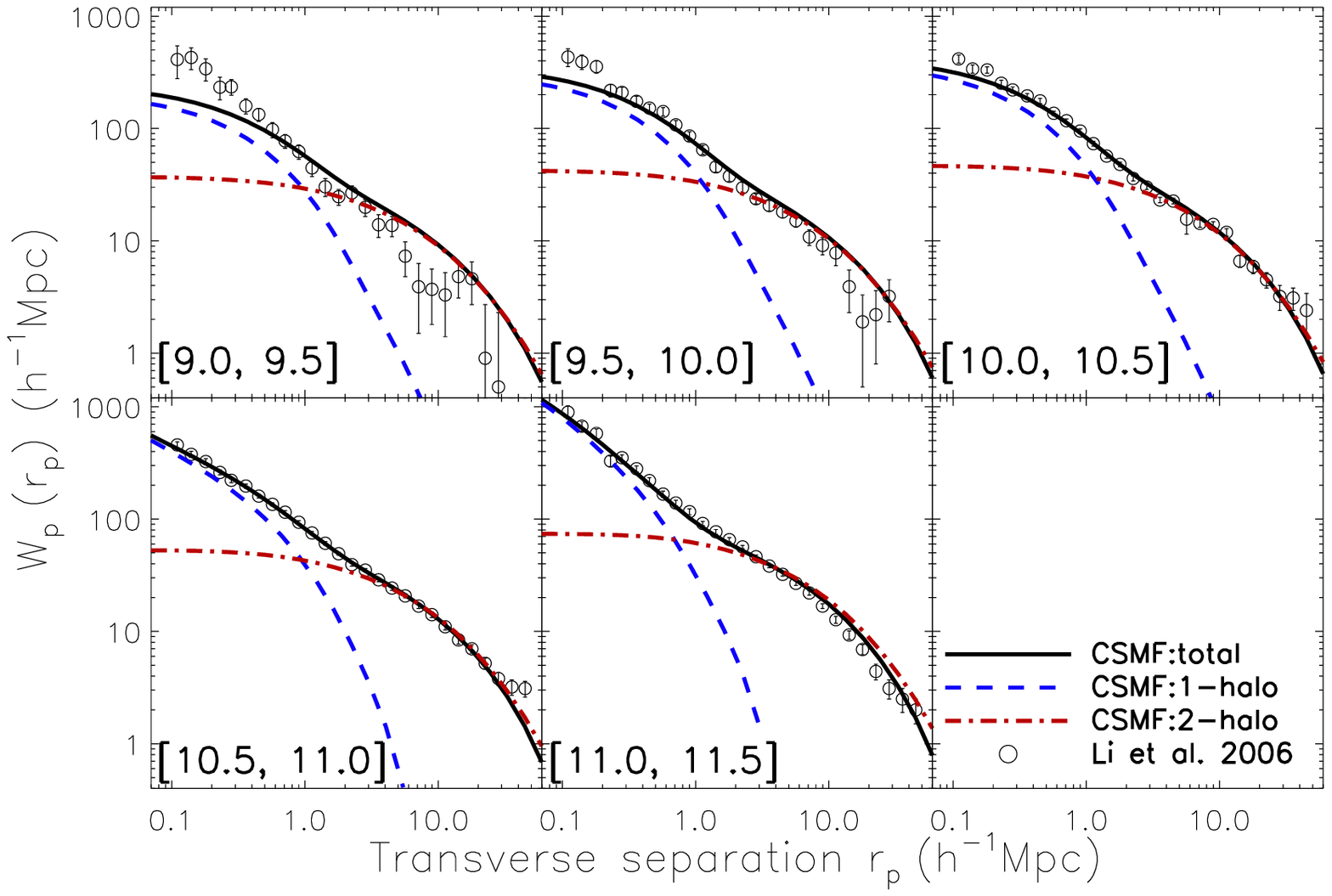}
\caption{Left: The measured SMF of the local Universe based on SDSS DR7 (Guo et al. 2010) compared with the SMF derived from our best-fit CSMF. Central galaxies (the red dot dashed line) dominate the SMF over the entire mass range probed. Right: The measured projected correlation functions of the SDSS galaxies in different stellar mass bins (Li et al. 2006) compared with the correlation functions derived from our best-fit CSMF. Note that the stellar mass bins shown in each panel are calculated with $h=0.7$. For example, the top left panel shows the projected correlation function of galaxies in the stellar mass bin $\log_{10} m_*(M_{\odot}) = [9.0, 9.5]$ assuming $h=0.7$.}
\label{fig:SMF}
\end{figure*}

We use the Le Phare code (Arnouts et al. 2002; Ilbert et al. 2006) and the Bruzual \& Charlot (2003) stellar population synthesis (SPS) models to derive stellar properties such as stellar mass and SFR.  We use the same parameters as in Ilbert et al. (2010) to generate the SED templates, e.g., a Chabrier initial mass function (IMF), two different metallicities (solar and sub-solar) and an exponentially declining star-formation history. Dust extinction is applied to the templates using the Calzetti et al. (2000) law.

We cross-match the 3.6 $\mu$m catalogue in each field with the 24 $\mu$m catalogue by taking the nearest match within 2". The SPIRE\footnote{The Spectral and Photometric Imaging Receiver (SPIRE; Griffin et al. 2010) is one of three scientific instruments on board Herschel (Pilbratt et al. 2010). It operates in three wavelength bands centred at 250, 350 and 500 $\mu$m.} fluxes of the 24 $\mu$m sources are obtained using a combination of linear inversion and model selection technique (Roseboom et al. 2010; Rosebomm et al. 2012).  With SPIRE, we are able to probe the rest-frame far-IR region to constrain the infrared luminosity $L_{\rm IR}$ (integrated from 8 to 1000 $\mu$m). Previous studies extrapolate $L_{\rm IR}$ from the 24 $\mu$m data and the resulting $L_{\rm IR}$ can be overestimated by a factor of five at $z>1.5$ (Papovich et al. 2007; Daddi et al. 2007; Murphy et al. 2009; Nordon et al. 2010; Elbaz et al. 2010). We use the Chary \& Elbaz (2001) templates to fit the infrared SEDs of galaxies observed at 24 $\mu$m and at least one SPIRE band to calculate $\psi_{\rm IR}=1.09\times10^{-10} \times L_{IR}$  (Kennicutt 1998). For galaxies not observed in any SPIRE band (around $70\%$ of the 3.6 $\mu$m - selected sample), we use $\psi_{\rm SED}$ derived from SED fitting to the UV to MIR photometric data.

\begin{table*}
\caption{A two-parameter fit of the form $\psi (M_\odot / \rm{yr}) = \alpha \times (m_*/M_\odot)^\beta$ to the stellar mass dependence of the average SFR for star-forming galaxies, averaged over ECDFS, COSMOS and EGS.}\label{table:mass_samples}
\begin{tabular}[pos]{lllllll}
\hline
\hline
Power-law fit         &           &  &  & Power-law slope fixed & & \\
$\Delta z$ & $\log_{10} \alpha$ [1/yr] & $\beta$ & $\chi^2/$dof &  $\log_{10} \alpha$ [1/yr] & $\beta$ & $\chi^2/$dof \\
\hline
\hline
$z=[0.2, 0.5] $        & $-2.08\pm2.26$      & $0.27\pm0.21$       & 0.17 & $-3.14\pm0.46$ & 0.37 & 0.20    \\
$z=[0.5, 0.8] $        & $-1.41\pm1.92$      & $0.25\pm0.18$       & 0.18 & $-2.73\pm0.37$ & 0.37 & 0.26    \\
$z=[0.8, 1.0] $        & $-1.52\pm2.04$      & $0.28\pm0.19$       & 0.30 & $-2.54\pm0.37$ & 0.37 & 0.34    \\
$z=[1.0, 1.3] $        & $-3.52\pm2.66$      & $0.49\pm0.25$       & 0.09 & $-2.20\pm0.40$ & 0.37& 0.14    \\
$z=[1.3, 1.6] $        & $-3.85\pm3.47$      & $0.56\pm0.32$       & 0.05 & $-1.83\pm0.36$ & 0.37 & 0.14    \\
$z=[1.6, 2.0] $        & $-1.86\pm2.65$      & $0.39\pm0.24$       & 0.03 & $-1.62\pm0.35$ & 0.37 & 0.03    \\
\hline
\end{tabular}
\end{table*}

%\footnote{For star-forming galaxies, the $m_*$ - $\psi$ relation is observed to be fairly tight (intrinsic scatter $\sim0.3$ dex) with a slope close to unity from $z\sim0$ to 2 (Elbaz et al. 2007; Daddi et al. 2007; Noeske et al. 2007; Rodighiero et al. 2010; Karim et al. 2011).}

In each field, we generate 10 Monte Carlo realisations of the original photo-$z$ catalogue using the redshift PDF of each galaxy and repeat the stellar mass and SFR calculation. In Fig.~\ref{fig:duty_cycle_conditional_highz}, we plot the conditional SFR distributions as a function of $m_*$ in 6 redshift bins, $z=[0.2, 0.5], [0.5, 0.8], [0.8, 1.0], [1.0, 1.3], [1.3, 1.6]$ and $[1.6, 2.0]$, averaged over all Monte Carlos realisations in COSMOS, ECDFS and EGS. The star-forming sequence\footnote{For star-forming galaxies, there exists a strong correlation between stellar mass $m_*$ and SFR $\psi$ (with an estimated intrinsic scatter $\sim0.3$ dex) from $z\sim0$ to 3 (e.g., Elbaz et al. 2007; Daddi et al. 2007; Noeske et al. 2007; Rodighiero et al. 2010; Karim et al. 2011).}  can be clearly seen and it evolves upwards roughly independently of $m_*$. The number of quiescent massive galaxies gradually builds up as redshift decreases. In each redshift bin, the conditional SFR distribution in a given stellar mass bin can be modelled as the sum of two Gaussian distributions which represent the star-forming and passive populations
\begin{equation}
\Phi(\psi|m_{\star}) = \Phi_{\rm star-forming}(\psi|m_{\star}) + \Phi_{\rm passive}(\psi|m_{\star}).
\end{equation}
In this paper, we define star-forming galaxies as those with SFR $\ge  \left<\psi\right>_{\rm star-forming}- 2\sigma_{\rm star-forming}$, where $ \left<\psi\right>_{\rm star-forming} $  and $\sigma_{\rm star-forming}$ is the mean SFR and standard deviation of the star-forming population respectively. The advantage of our definition of star-forming galaxies is that it naturally takes into account the fact that the SFR of a star-forming galaxy increases with increasing stellar mass and increasing redshift (as shown in Fig. 1). In Fig.~\ref{fig:duty_cycle_conditional_highz_fit}, we plot the redshift evolution of the average SFR as a function of stellar mass for star-forming galaxies and the best-fit power-law to points above the stellar mass completeness limit in each redshift slice (see Table 1). The best-fit parameters in the power-law fitting of the $m_{\star}$ - $\psi$ relation are listed in Table 2.

\begin{figure}\centering
\includegraphics[height=2.7in,width=3.5in]{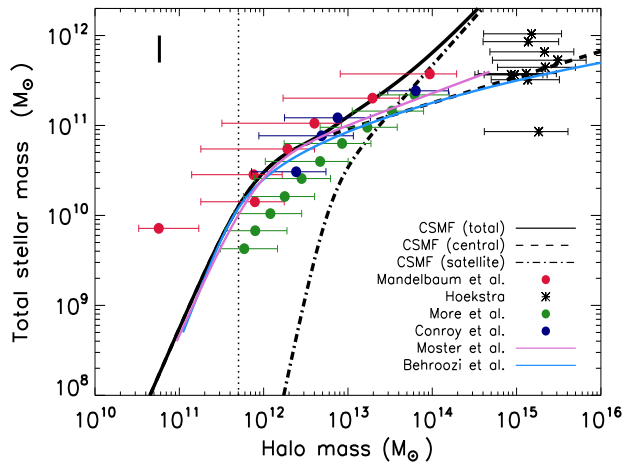}
\caption{The predicted average total stellar mass as a function of $M_{\rm h}$ derived from the best-fit CSMF of the local Universe. The predicted average stellar mass of central galaxies agrees reasonably well with results from galaxy-galaxy lensing (Mandelbam et al. 2006; Hoekstra 2007), satellite kinematics (Conroy et al. 2007; More et al. 2011), and other empirical models of the stellar-to-halo mass relation (Behroozi et al. 2010; Moster et al. 2010). The vertical bar on the left indicates the typical error in $m_*$. The vertical dotted line marks the characteristic $M_{\rm h}$ in the $m_*$ - $M_{\rm h}$ relation for central galaxies. }
\label{fig:sm_hm}
\end{figure}

\section{EHM: 1. Connecting stellar mass with halo mass}

\subsection{The stellar-to-halo mass relation at $z\sim0$}

%The effective bias for the different stellar mass bins are:
 %     1.13698      1.23417      1.30527      1.41662      1.73130
%The effective halo mass for the different stellar mass bins are:
 % 3.41861e+13  4.02499e+13  4.26527e+13  4.49447e+13  6.43580e+13

We choose the CSMF, $\Phi(m_*|M_{\rm h})$, which specifies the number of galaxies of stellar mass $m_*$ that reside in a halo of mass $M_{\rm h}$, to describe the stellar-to-halo mass relation. Details of the parametrisation of the CSMF and the fitting process to the observed spatial density and clustering of galaxies can be found in Appendix~\ref{first appendix}. The left panel in Fig.~\ref{fig:SMF} compares the measured SMF of the local Universe (Guo et al. 2010) with the best-fit SMF from our CSMF. Note that in comparing to the observed SMF, the predicted SMF from the CSMF (the black dashed line in the left panel of Fig.~\ref{fig:SMF}) has been convolved with a log-normal distribution with its width set to 0.1 dex (Li \& White 2009) to account for statistical errors in the observational estimate of stellar mass. It is clear that central galaxies dominate the SMF over the entire mass range probed. At $m_*>10^8 M_{\odot}$, satellite galaxies make up $18\%$ of the entire population. The fraction of satellite galaxies decreases rapidly with increasing stellar mass at the high mass end. At $m_*>5\times10^{10} M_{\odot}$, satellite galaxies account for $8\%$ of the entire population while at $m_*>10^{11} M_{\odot}$, the fraction of satellites is $<1\%$.  The projected correlation functions in five stellar mass bins (Li et al. 2006) are compared with the best-fit from our CSMF in the right panel in Fig.~\ref{fig:SMF}. The 1-halo term (due to galaxy pairs residing in the same halos) dominates the clustering signal on small scales and the 2-halo term (due to galaxy pairs in separate halos) dominates the clustering signal on large scales. The transition between the 1-halo and 2-halo term is at $\sim1/h$ Mpc in all stellar mass bins. The large-scale 2-halo term (proportional to the linear bias factor) increases with $m_*$ indicating more massive galaxies reside in more massive halos. In the two lowest stellar mass bins, the predicted clustering signal lies below the measured on the smallest scales ($\lesssim 0.2 h^{-1}$ Mpc). It can not be due to our particular choice of the galaxy density profile inside a dark matter halo because we do not see the same effect in the three more massive mass bins. A full investigation of the cause is deferred until the full covariance matrix of the correlation function is available.

\begin{table}
\caption{Volume-limited and stellar-mass-selected subsamples in COSMOS and EGS used to calculate correlation functions. The columns are sample name, redshift range, number of galaxies and stellar mass range. Note that the number of galaxies in each sample varies slightly in different Monte Carlo realisations.}\label{table:mass_samples}
\begin{tabular}[pos]{llll}
\hline
\hline
Sample           & $z$  range         & $N_{\rm gal}$ & $\log_{10} m_*$  ($M_{\odot}$)\\
\hline
\hline
z1M1 (COSMOS)         & $z=[0.2, 0.5]$       & 2117          & [9.8, 10.1]     \\
z1M2 (COSMOS)        & $z=[0.2, 0.5] $      & 2025           & [10.1,10.4]  \\
z1M3 (COSMOS)         & $z=[0.2, 0.5] $      & 2175          & $>10.4$  \\ 
z2M1 (COSMOS)         & $z=[0.5, 0.8] $      & 2311          & [10.1, 10.4] \\
z2M2 (COSMOS)        & $z=[0.5, 0.8] $      & 2641          & [10.4, 10.6] \\
z2M3 (COSMOS)         & $z=[0.5, 0.8]$       & 2369          & $>10.6$  \\ 
z3M1 (COSMOS)         & $z=[0.8, 1.0] $      & 4821         & [10.1, 10.5]\\  
z3M2 (COSMOS)        & $z=[0.8, 1.0] $      & 5051          & $>10.5$\\
z4M1 (COSMOS)       & $z=[1.0, 1.3] $      & 4111          & [10.2, 10.6]\\
z4M2 (COSMOS)         & $z=[1.0, 1.3] $      & 3950          & $>10.6$\\
z5M1 (COSMOS)        & $z=[1.3, 1.6] $      & 3867          & $>10.4$ \\
z6M1 (COSMOS)        & $z=[1.6, 2.0] $      & 2425          & $>10.7$ \\ 
\hline
z1M1 (EGS)         & $z=[0.2, 0.5]$       & 2064          & $>$9.9     \\
z2M1 (EGS)         & $z=[0.5, 0.8] $      & 2186          & [10.0, 10.2] \\
z2M2 (EGS)         & $z=[0.5, 0.8] $      & 2045          & $>$10.2 \\
z3M1 (EGS)         & $z=[0.8, 1.0] $      & 2650          & $>$10.1\\  
z4M1 (EGS)        & $z=[1.0, 1.3]  $     & 2965          & $>$10.1\\
z5M1 (EGS)         & $z=[1.3, 1.6] $      & 2333          & $>$10.3 \\
z6M1 (EGS)         & $z=[1.6, 2.0] $      & 1731          & $>$10.6 \\ 
\hline
\end{tabular}
\end{table}

With the parameters in the CSMF tuned by the galaxy abundance and clustering data, we can now predict the average total stellar mass as a function of halo mass can be calculated from the best-fit CSMF,
%\begin{equation}
%\left<m_{\star}\right> = \int m_{\star} \times \Phi(m_{\star}|M_{\rm h}) dm_{\star},
% \end{equation} 
\begin{eqnarray}
\left<m_{\star}\right>_{\rm total} &=& \int m_{\star} \times \Phi(m_{\star}|M_{\rm h}) dm_{\star} \nonumber \\
                                     &=& \int m_{\star} \times [\Phi_{\rm cen}(m_{\star}|M_{\rm h}) + \Phi_{\rm sat}(m_{\star}|M_{\rm h})] dm_{\star} \nonumber \\
                                     	& = & \left<m_{\star}\right>_{\rm cen} + \left<m_{\star}\right>_{\rm sat},
 \end{eqnarray}  
which is plotted in Fig.~\ref{fig:sm_hm}. The average stellar mass of the central galaxies as a function of halo mass from our CSMF model agrees reasonably well with constraints from galaxy-galaxy lensing (Mandelbaum et al. 2006; Hoekstra 2007), satellite dynamics (Conroy et al 2007.; More et al. 2011) and galaxy group catalogues (Yang et al. 2009). Our result on the stellar-to-halo mass relation for central galaxies also agrees well with other empirical models, i.e. Moster et al. (2010) and Behroozi et al (2010)\footnote{Moster et al. (2010) uses an almost identical CSMF formalism to what is used in this paper. Essentially, it is a double power-law connected at some characteristic mass scale. Behroozi et al (2010) uses a different CSMF formalisation. The main difference is that for high-mass galaxies, the $m_*$ - $M_{\rm h}$ relation asymptotes to a sub-exponential function instead of a power-law.}. Both Moster et al. (2010) and Behroozi et al. (2010) fit to the observed SMF only. The good agreement between different empirical models indicates that an accurate SMF is the most important constraint in determining the statistical relation between $m_{\star}$ and $M_{\rm h}$. In our CSMF model, the average $m_*$ of the central galaxies grows roughly as $M_{\rm h}^{1.16}$ at the low-mass end and as $M_{\rm h}^{0.71}$ at the high-mass end. The characteristic halo mass for central galaxies in our model, which is where the low- and high-mass power-laws meet, is $5\times10^{11}M_{\odot}$. The corresponding stellar mass at the characteristic halo mass is $\sim10^{10}M_{\odot}$, which is where local galaxies are found to divide into two distinct families with less massive galaxies showing younger stellar populations, optically blue colours and disk-like morphologies, and more massive galaxies exhibiting older stellar populations, optically red colours, and more bulge-like morphology (Kauffmann et al. 2003). Therefore, the different stellar mass build-up history, indicated by the different $m_*$ - $M_{\rm h}$ relation below and above $M_{\rm h}=5\times10^{11}M_{\odot}$, may explain the observed division in galaxy properties below and above $m_{\star}\sim10^{10}M_{\odot}$.

\subsection{The stellar-to-halo mass relation at high $z$}

\begin{figure*}\centering
\includegraphics[height=3.in,width=3.45in]{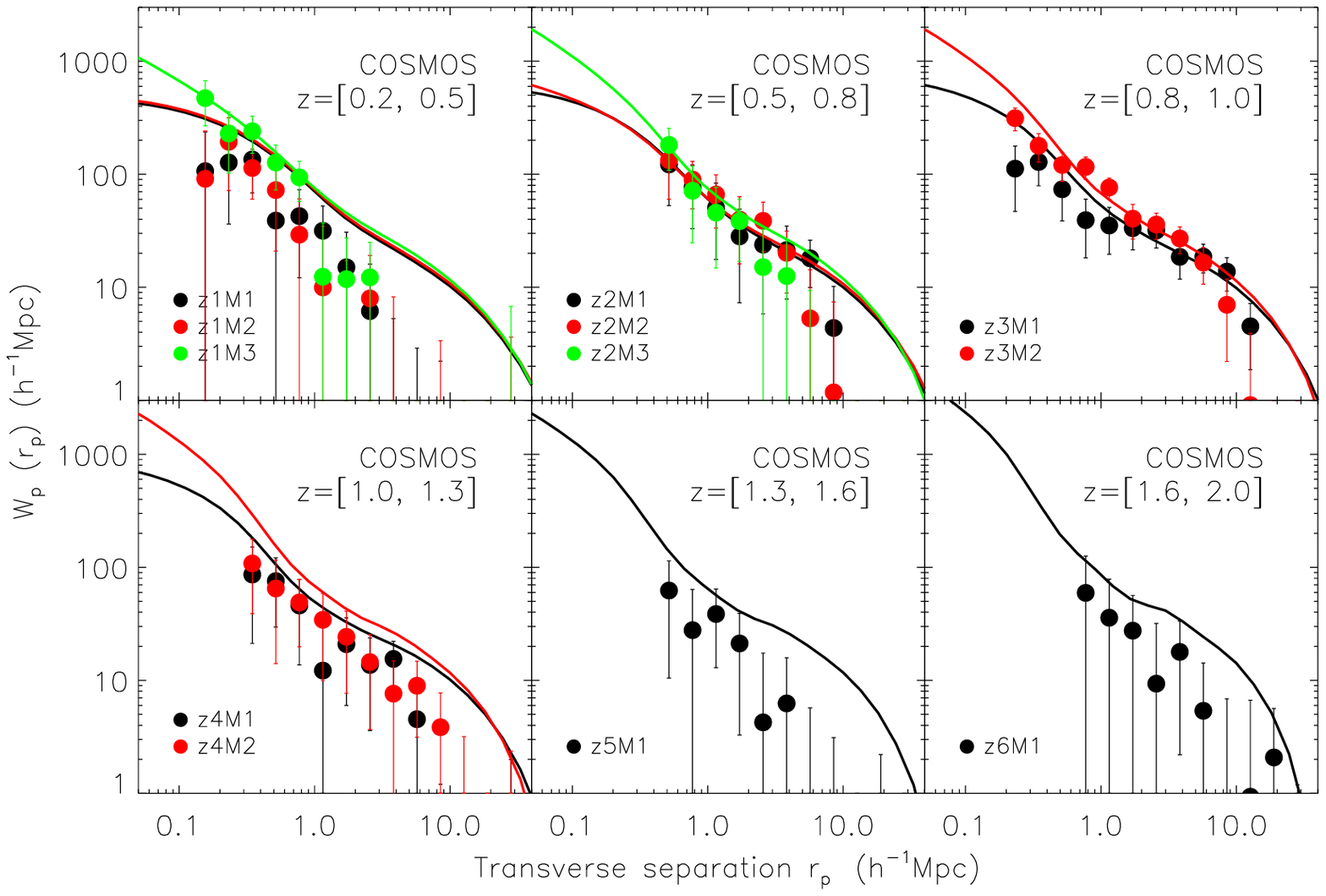}
\includegraphics[height=3.in,width=3.45in]{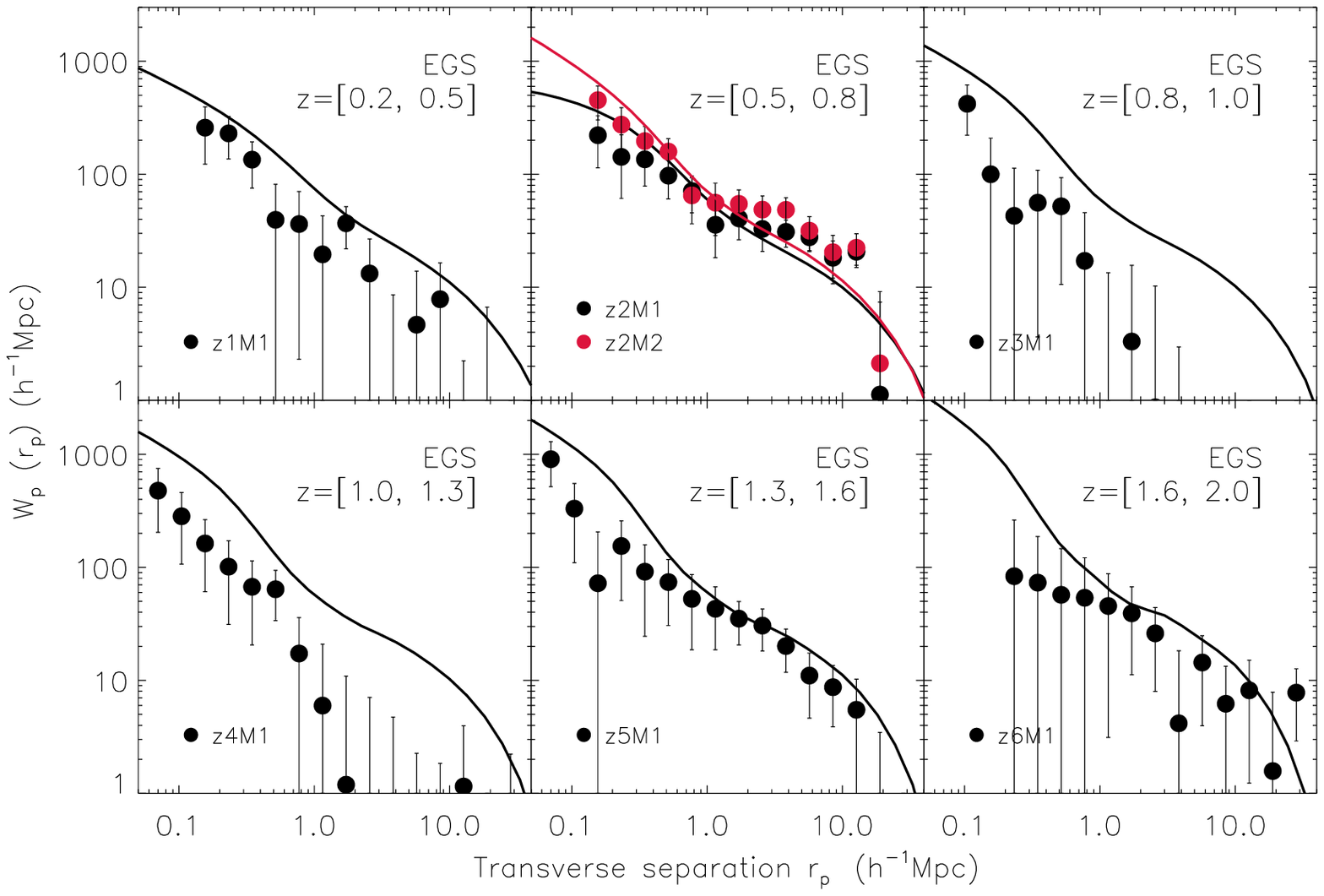}
\caption{The projected correlation functions of stellar-mass-limited subamples listed in Table 3 in six redshift bins $z1=[0.2, 0.5]$, $z2=[0.5, 0.8]$, $z3=[0.8, 1.0]$, $z4=[1.0, 1.3]$, $z5=[1.3, 1.6]$ and $z6=[1.6, 2.0]$. The solid lines are the predicted correlation function from our best-fit CSMF. Error bars include both the bootstrapping error and the photometric redshift error. In redshift bins where multiple stellar-mass-limited subsamples exist, more massive galaxies seem to show a higher clustering amplitude than less massive galaxies.}
\label{fig:wp_sm_cosmos}
\end{figure*}

\begin{figure*}\centering
\includegraphics[height=1.6in,width=4.4in]{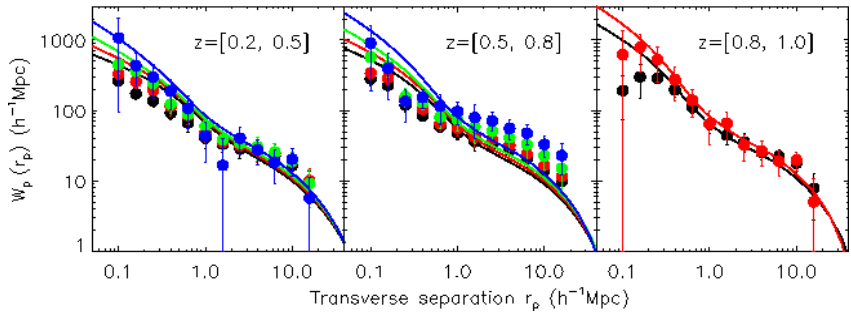}
\caption{The projected correlation functions of stellar-mass-limited samples (black points: galaxies with $\log (m_*/M_\odot) \ge 9.0$; red points: $\log (m_*/M_\odot) \ge 9.5$; green points: $\log (m_*/M_\odot) \ge 10.0$; blue points: $\log (m_*/M_\odot) \ge 10.5$) in three redshift bins $z1=[0.2, 0.5]$, $z2=[0.5, 0.8]$ and $z3=[0.8, 1.0]$ from the zCOSMOS 10K sample (Meneux et al 2009). The solid lines are the predicted correlation function from our best-fit CSMF. The flat shape in the measured zCOMOS correlation functions in the middle panel has been explained by an overabundance of high-density regions (de la Torre et al. 2010).}
\label{fig:wp_sm_cosmos_meneux}
\end{figure*}

In Table 3, we list a series of volume- and stellar-mass-limited subsamples in six redshift bins in COSMOS and EGS. The projected correlation function for each subsample in COSMOS and EGS is plotted in Fig.~\ref{fig:wp_sm_cosmos}. More details on how projected correlation functions are calculated can be found in Appendix B. In redshift bins where multiple stellar-mass-limited subsamples exist, it seems that more massive galaxies generally show stronger clustering although the large errors prevent any firm conclusions to be drawn. This is consistent with Meneux et al. (2009) who studied the clustering dependence on $m_*$ in the redshift bin $z=[0.2, 1]$ using the first 10K redshifts from the zCOSMOS survey and found a mild dependence on $m_*$ especially on small scales (see Fig.~\ref{fig:wp_sm_cosmos_meneux}). 

\begin{figure*}\centering
\includegraphics[height=2.6in,width=4.6in]{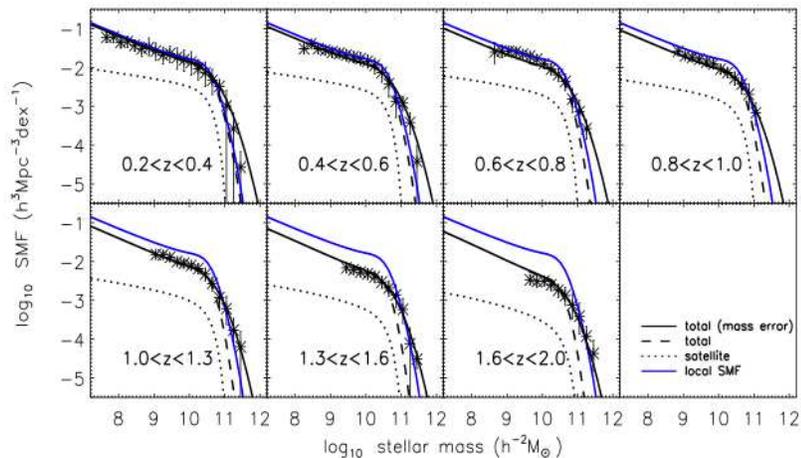}
\caption{The measured SMFs in different redshift bins from P{\'e}rez-Gonz{\'a}lez et al. (2008). The redshift range is indicated in each panel. The dashed black line in each panel is the underlying SMF predicted from the best-fit CSMF. The solid black line in each panel is the convolution of the dashed black line with a log-normal distribution which represents the statistical error in the observational estimate of stellar mass. The blue line is the present-day SMF (i.e. the black dashed line in the left panel of Fig.~\ref{fig:SMF}).}
\label{fig:SMF_cosmos}
\end{figure*}

We derive the $m_*$ - $M_{\rm h}$ relation for the local Universe by fitting to both the spatial density and clustering of galaxies. At high $z$, however, we will only use the SMFs and not the correlation functions presented above. This is because the correlation function is extremely sensitive to cosmic variance. A large difference in the correlation functions between COSMOS and VVDS was reported in Meneux et al. (2009). Also, the flat shape in the measured zCOMOS correlation functions (shown in Fig.~\ref{fig:wp_sm_cosmos_meneux}) over the redshift range $z=[0.6, 1.0]$ has been attributed to an overabundance of high-density regions (de la Torre et al. 2010).
We show the measured SMFs in P{\'e}rez-Gonz{\'a}lez et al. (2008) based on a combined sample of 3.6 and 4.5 $\mu$m selected sources in the HDF-N, the CDF-S and the Lockman Hole and the best-fit from our CSMF model in Fig.~\ref{fig:SMF_cosmos}. Note that in comparing to the observed SMF, the predicted SMF (i.e. the intrinsic SMF) from the best-fit CSMF model (the black dashed line in Fig.~\ref{fig:SMF_cosmos}) has been convolved with a log-normal distribution with its width set to 0.3 dex (P{\'e}rez-Gonz{\'a}lez et al. 2008) to account for statistical errors in the observational estimate of stellar mass. The SMF increases over time but mostly in low-mass systems. The contribution from satellites also grows over time. In Fig.~\ref{fig:wp_sm_cosmos}, the projected correlation functions in COSMOS and EGS are compared with the predicted correlation functions from our best-fit CSMF. There is a relatively good overall agreement between the two. On large scales, the measured correlation function falls under the predicted curve, which is due to integral constraint. If the galaxy number density fluctuations in the probed volume are smaller than the average over a cosmologically representative volume, then the measured correlation function will be biased low by a constant, which is equal to the fractional variance of the number counts in cells. This effect is significant if the survey field is small.  In Fig.~\ref{fig:wp_sm_cosmos_meneux}, we compare the projected correlation functions of stellar-mass-limited samples from the zCOSMOS 10K sample (Meneux et al. 2009) with the predicted clustering from our model and again find relatively good overall agreement.

\begin{figure*}\centering
\includegraphics[height=2.6in,width=3.46in]{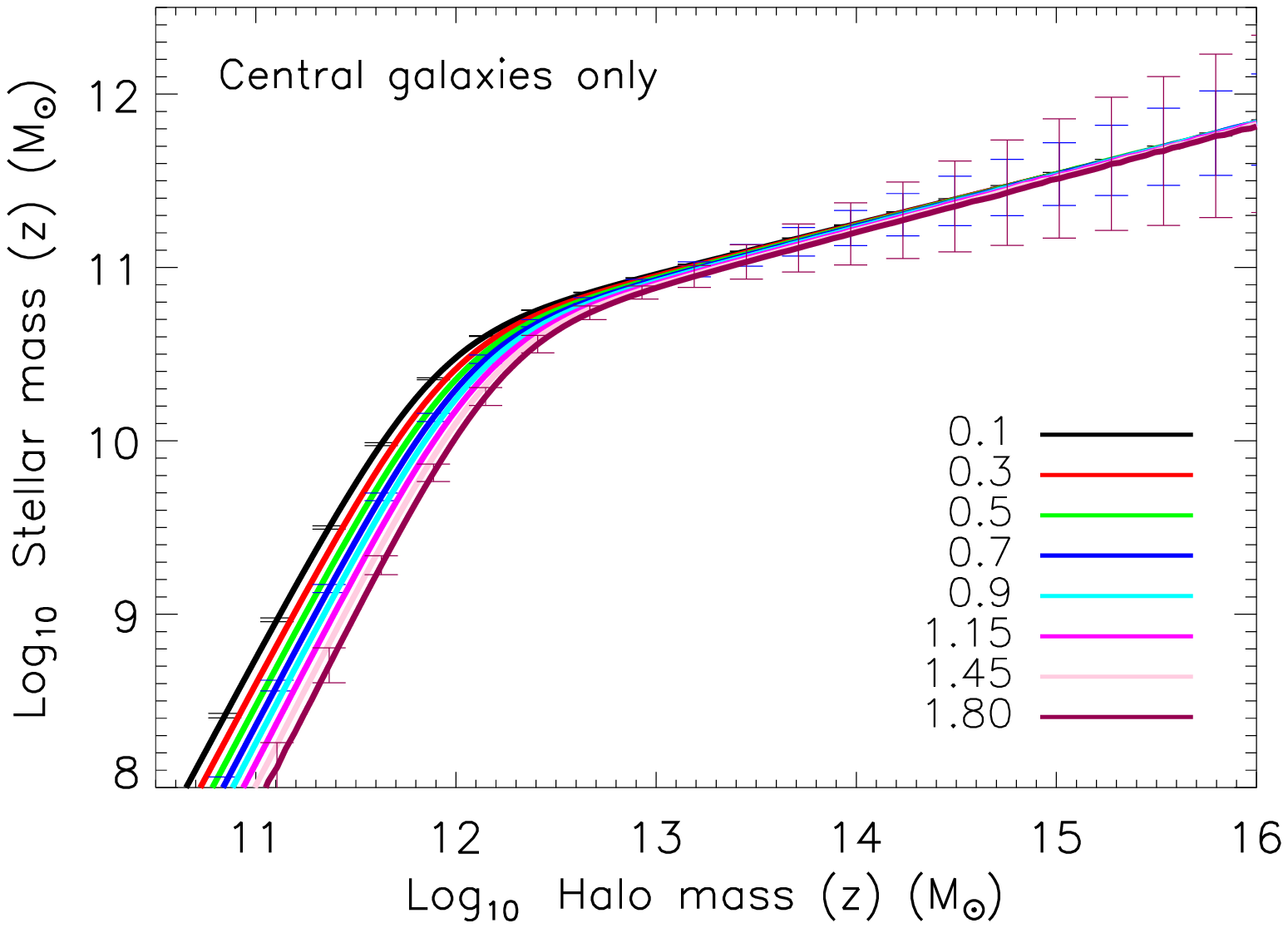}
\includegraphics[height=2.6in,width=3.46in]{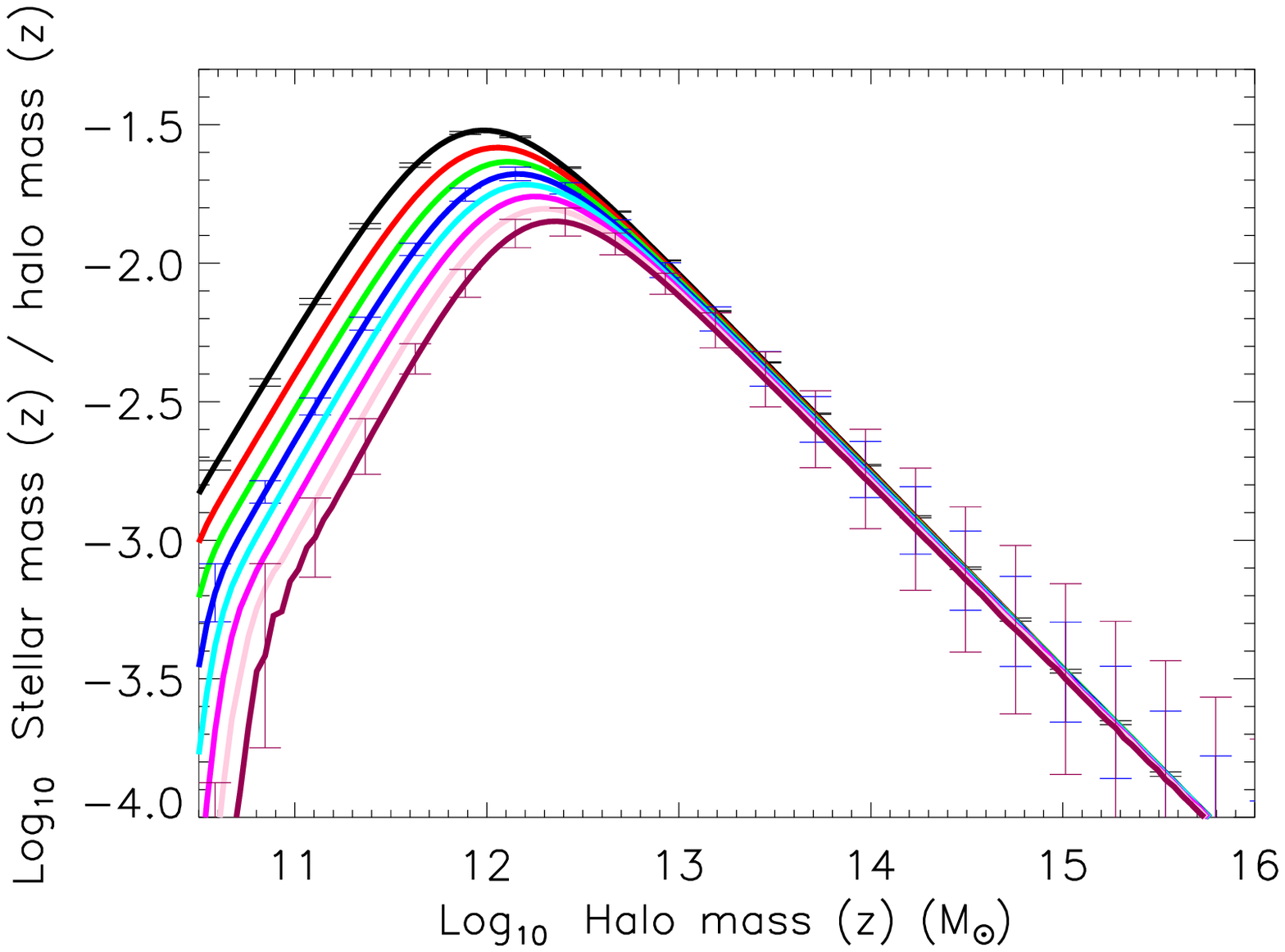}
\caption{Left: The predicted average stellar mass of central galaxies as a function of halo mass from the best-fit CSMF. For clarity, we only plot errors on a few selected redshift slices and halo mass bins. Different lines are colour-coded by redshift as indicated in the panel.  The characteristic halo mass scale for central galaxies increases with increasing redshift. Right: The average stellar-to-halo mass ratio for the central galaxies versus the host halo mass predicted from the best-fit CSMF. It is clear that the star-formation efficiency is low in both low-mass and high-mass halos and the peak in the stellar-to-halo mass ratio shifts to lower mass halos over time.}
\label{fig:Mstar_Mhalo_z}
\end{figure*}

We plot the average $m_*$ - $M_{\rm h}$ relation as a function of $M_{\rm h}$ for central galaxies  in the left panel in Fig.~\ref{fig:Mstar_Mhalo_z}. The characteristic halo mass scale in the $m_*$ - $M_{\rm h}$ relation for central galaxies has increased with increasing redshift, changing from $\sim5.0\times10^{11}M_{\odot}$ at $z=0.1$ to $\sim1.1\times10^{12}M_{\odot}$ at $z=1.8$. In the right panel of  Fig.~\ref{fig:Mstar_Mhalo_z}, we plot the $m_*$-to-$M_{\rm h}$ (a measure of the integrated star-formation efficiency) as a function of $M_{\rm h}$ for central galaxies. It is clear that the integrated star-formation efficiency is low in both low-mass and high-mass halos in all redshift slices at $0<z<2$. In low-mass halos, star-formation efficiency is suppressed possibly due to supernova feedback which can re-heat the interstellar stellar medium, heat gas in the dark matter halo or even  eject gas altogether (Springel \& Hernquist 2003; Brooks et al. 2007; Ceverino \& Klypin 2009). In high-mass halos, star-formation efficiency is also suppressed possibly due to gravitational heating (Khochfar \& Ostriker 2008; Dekel \& Birnboim 2008) and/or feedback from AGN which transfers energy to the halo gas (Croton et al. 2006; Bower et al 2006; Monaco et al. 2007). The peak of the average stellar-to-halo mass ratio for central galaxies has shifted towards lower mass halos over time. %At $z\sim2$, the peak of the accumulated stellar mass growth occurs at $M_{\rm h}\sim10^{12.5}M_{\odot}$. At $z\sim0$, the peak of the accumulated stellar mass growth occurs at $M_{\rm h}\sim10^{12.3}M_{\odot}$. 

\begin{figure*}\centering
\includegraphics[height=2.6in,width=3.46in]{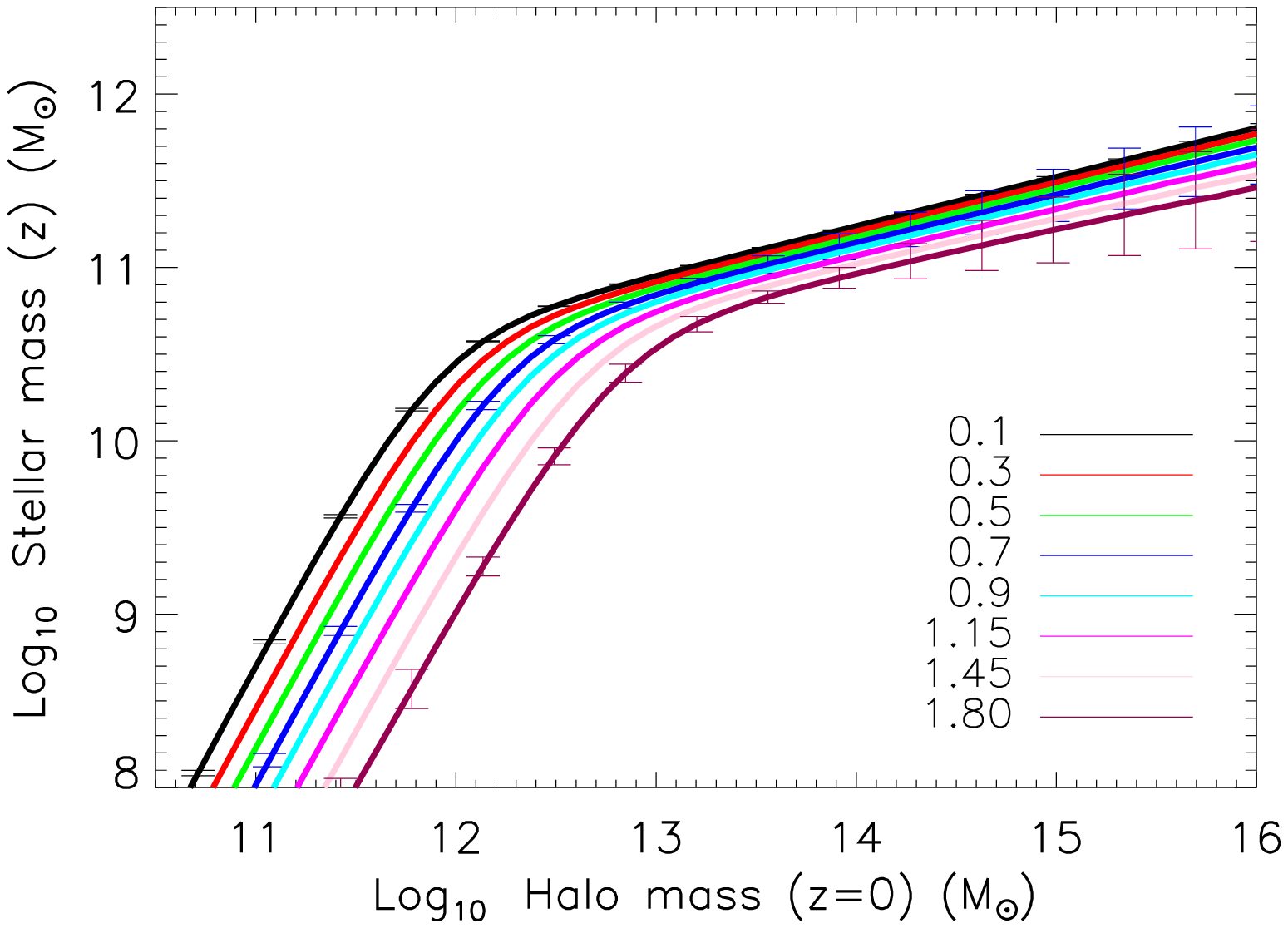}
\includegraphics[height=2.6in,width=3.46in]{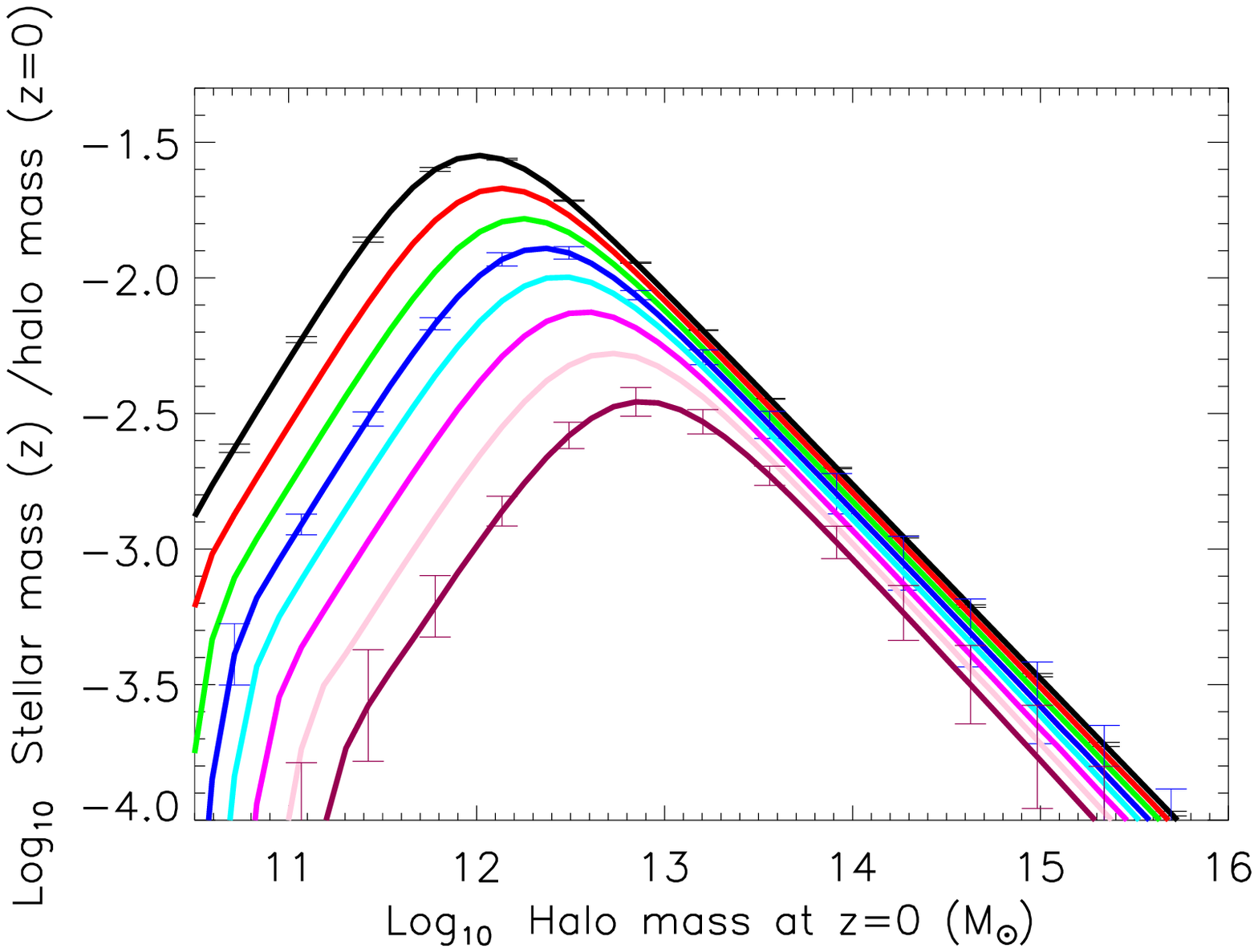}
\caption{Left: The predicted average stellar mass of the central galaxies as a function of halo mass at the present day. The halo mass is evolved to $z=0$ using the halo mass accretion history from Fakhouri et al. (2010). The evolution of the stellar content as a function of halo mass. Different lines are colour-coded by redshift as indicated in the panel.  Right: The average stellar-to-halo mass ratio versus halo mass at the present day. The build-up of stellar mass happened early on in massive halos.}
\label{fig:Mstar_Mhalo_z=0}
\end{figure*}

In Fig.~\ref{fig:Mstar_Mhalo_z=0}, we plot the stellar mass build-up history as a function of halo mass at the present day by evolving $M_{\rm h}$ at a particular redshift to $M_{\rm h}$ at $z=0$ using the halo mass accretion rate from Fakhouri et al. (2010),
\begin{equation}
\left<\frac{dM_{\rm h}}{dt}\right> = 46.1 \left(\frac{M_{\rm h}}{10^{12}}\right)^{1.1} (1+1.11z) \sqrt{\Omega_m(1+z)^3 + \Omega_{\Lambda}}.
\end{equation}
So we can trace the evolution of the stellar content in the same halo along any vertical line in Fig.~\ref{fig:Mstar_Mhalo_z=0}. It is clear that the stellar mass assembly happened much earlier in massive halos than in less massive halos. In halos more massive than $10^{13} M_{\odot}$ (the present-day value), the stellar mass of the central galaxies has increased by at most a factor of a few. But in less massive halos, the stellar mass of the central galaxies has grown by an order of magnitude or more. This is consistent with the downsizing scenario of galaxy formation.

%\begin{figure*}\centering
%\includegraphics[height=2.9in,width=3.46in]{Mstar_Mhalo_highz_part2_compare_behroozi.eps}
%\includegraphics[height=2.9in,width=3.46in]{Mstar_Mhalo_highz_part2_compare_leathaud.eps}
%\includegraphics[height=2.9in,width=3.46in]{Mstar_Mhalo_highz_part2_compare_moster.eps}
%\caption{Top: The average stellar-to-halo mass ratio versus halo mass. The peak in the stellar-to-halo mass ratio shifts to lower mass halos over time. Different lines are colour-coded by redshift as indicated in the panel. Bottom: The evolution of the stellar content as a function of halo mass. The build-up of stellar mass happened early on in massive halos. Top: The average stellar-to-halo mass ratio versus halo mass redshift evolved to $z=0.1$ using the halo mass accretion history from Fakhouri et al. (2010). The peak in the stellar-to-halo mass ratio shifts to lower mass halos over time. Different lines are colour-coded by redshift as indicated in the panel. Bottom: The evolution of the stellar content as a function of halo mass redshift evolved to $z=0.1$. The build-up of stellar mass happened early on in massive halos.}
%\label{fig:Mstar_Mhalo_z}
%\end{figure*}

\section{EHM: 2. connecting stellar mass, SFR and halo mass}

Now we can extend the CSMF to the 2-D distribution $\Phi(\psi, m_{\star}|M_{\rm h})$, which specifies the number of galaxies as a function of $m_*$ and $\psi$ at fixed $M_{\rm h}$. Using conditional probability theory, one can show that
\begin{equation}
\Phi(\psi, m_{\star}|M_{\rm h}) = \Phi(m_{\star}|M_{\rm h}) \times \Phi(\psi|m_{\star}, M_{\rm h}).
\end{equation}
 If the distribution of SFR is only dependent on $m_*$ and at most weakly dependent on $M_{\rm h}$, then one can assume
\begin{equation}
\Phi(\psi, m_{\star}|M_{\rm h})  \approx \Phi(m_{\star}|M_{\rm h}) \times\Phi(\psi|m_{\star}).
\end{equation}
We will refer to this simplification as  {\bf Scenario A}.

%which satisfy $\log_{10} \psi / m_* \leq -9 - 0.2 \log [m_*/h^{-2} M_{\odot}]$)

However, it is important to realise that the distribution of SFR at fixed $m_*$ may be different in halos of different masses, which is a measure of the Mpc-scale environment. 
Using group catalogues constructed from the SDSS DR5 (Yang et al. 2007), Kimm et al. (2009) studied the fraction of passive galaxies, $f_{\rm passive}$, as a function of $m_*$ and $M_{\rm h}$. Within the error bars, it is difficult to tell whether $f_{\rm passive}$ at fixed $m_*$ has any significant dependence on $M_{\rm h}$. However, Peng et al. (2010) using both the SDSS and zCOSMOS dataset found that the SFR of star-forming galaxies at fixed $m_*$ is completely independent of environment (measured by the 5th nearest neighbour density estimator)\footnote{Since the SFR distribution at a given stellar mass is independent of environment for star-forming galaxies, we will only need to use {\bf Scenario A} to connect SFR with halo mass for the star-forming galaxy population.}, but $f_{\rm passive}$ depends on environment even at fixed $m_*$. 
Therefore, in this paper, we adopt a second scenario in building the 2-D distribution in the ($\psi, m_*$) plane as a function on halo mass.
We assume that the fraction of passive galaxies at fixed $m_*$ has a power-dependence on $M_{\rm h}$, i.e. $f_{\rm passive}(M_{\rm h}|m_{\star}) \propto M_{\rm h}^{\eta(m_{\star})}$. Furthermore we assume that all galaxies are passive in very massive halos  (corresponding to the most massive rich clusters), i.e. $f_{\rm passive} = 1$ at $M_{\rm h} =10^{15} M_{\odot}$. Since we know the overall $f_{\rm passive}$ in a given stellar mass bin, we can work out the power-law dependence $\eta(m_{\star})$. Under this assumption, the SFR distribution at fixed $m_{\star}$ and $M_{\rm h}$ can be derived from the SFR distribution at fixed $m_{\star}$ but with $f_{\rm passive}$ modulated by halo mass, i.e.
\begin{equation}
\Phi(\psi|m_{\star}, M_{\rm h}) \approx \Phi(\psi|m_{\star}) f_{\rm passive} (M_{\rm h}| m_{\star}).
\end{equation}
We will refer to this simplification as  {\bf Scenario B}.

\begin{figure*}\centering
\includegraphics[height=2.9in,width=3.45in]{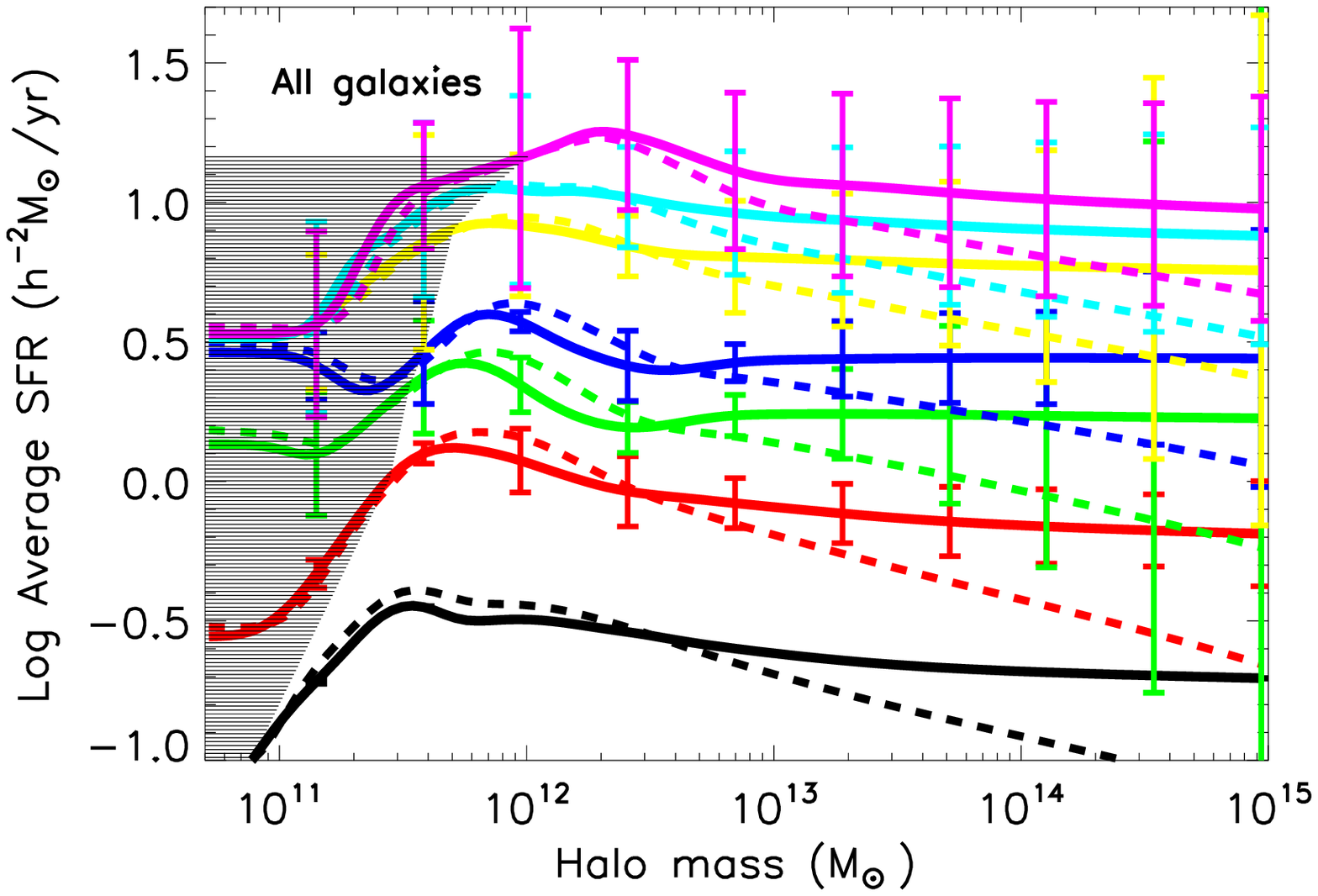}
\includegraphics[height=2.9in,width=3.45in]{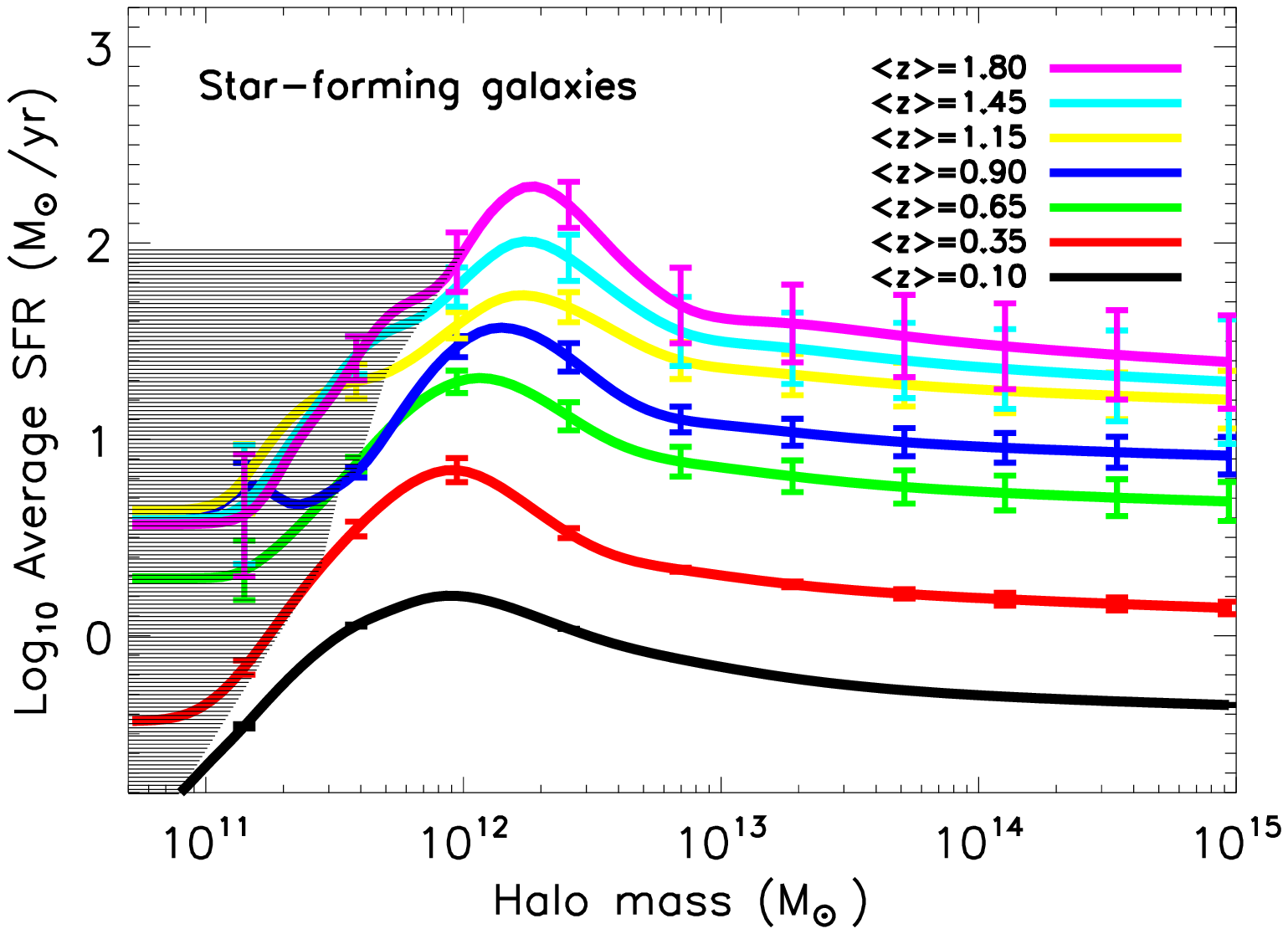}
%\includegraphics[height=2.9in,width=3.45in]{SFR_halomass_highz_halomassz0_sfgalaxiesonly.eps}
%\caption{Left: The average SFR as a function of $M_{\rm h}$. Different lines are colour-coded by redshift as indicated. The solid/dashed lines correspond to the $\psi$ - $M_{\rm h}$ relation derived from {\bf Scenario A} / {\bf B}.  The dark red regions indicate the $M_{\rm h}$ range where we are not able to derive reliable constraints on the $\psi$ - $M_{\rm h}$ relation due to the increasingly limited $m_*$ range probed towards higher $z$. Right: Similar to the left panel but the x-axis is $M_{\rm h}$ evolved to $z=0.1$ using the halo accretion history from Fakhouri et al. (2010). Along any vertical line, we can trace the evolution of the SFR in the same halo. The dark grey / light grey / white region indicates $M_{\rm h}$  range typically associated with cluster / group /  field environment.}
\caption{Left: The average SFR as a function of $M_{\rm h}$. Error bars include the uncertainty in the parametrised stellar-to-halo mass relation, the field-to-field variation in the SFR distribution as a function of stellar mass and the photometric redshift error. At z=0.1, the errors on the SFR - $M_{\rm h}$ relation is very small because the error bars only include the uncertainty in the parametrised stellar-to-halo mass relation. Different lines are colour-coded by redshift as indicated. The solid/dashed lines correspond to the $\psi$ - $M_{\rm h}$ relation derived from {\bf Scenario A} / {\bf B}. The hatched regions indicate the $M_{\rm h}$ range where we are not able to derive reliable constraints on the $\psi$ - $M_{\rm h}$ relation due to the increasingly limited $m_*$ range probed towards higher $z$. Right: Similar to the left panel but for star-forming galaxies only. Since the SFR distribution at a given stellar mass for star-forming galaxies is independent of environment, only {\bf Scenario A} is plotted.}
\label{fig:SFR_HM}
\end{figure*}

In Fig.~\ref{fig:SFR_HM}, we plot the average total SFR as a function of $M_{\rm h}$ at various redshifts. The left panel is for all galaxies and the right panel is for star-forming galaxies only (as defined in Section 2.4). Error bars include the uncertainty in the parametrised stellar-to-halo mass relation, the field-to-field variation in the SFR distribution as a function of stellar mass and the photometric redshift error. At $z=0.1$, the errors on the SFR - $M_{\rm h}$ relation is very small because the error bars only include the uncertainty in the parametrised stellar-to-halo mass relation. The average SFR is higher/lower in less/more massive halos in {\bf Scenario B} than in {\bf Scenario A}. This is because {\bf Scenario B} assumes that the $f_{\rm passive}$ increases with increasing $M_{\rm h}$. However, the difference in the SFR as a function of $M_{\rm h}$ between the two scenarios is small and does not affect the qualitative conclusions drawn in this paper. We can see that the intensity of star-forming activity in halos in the probed mass range has steadily decreased as a function of time, dropping by over one order of magnitude from $z\sim2$ to $z\sim0$. The peak in SFR shifts from $M_{\rm h}$ just over $10^{12} M_{\odot}$ at $z\sim2$ to just below $10^{12} M_{\odot}$ at $z\sim0.1$, in qualitative agreement with Fig.~\ref{fig:Mstar_Mhalo_z} where the peak of the stellar-to-halo mass ratio (a measure of the integrated star-formation efficiency) is shown to shift towards lower mass halos over time. At a given redshift, halos in the mass range between a few times $10^{11} M_{\odot}$ and a few times $10^{12}M_{\odot}$ are the most efficient at hosting star formation. Again, this is consistent with Fig.~\ref{fig:Mstar_Mhalo_z} which shows the integrated star-formation efficiency is low in both low- and high-mass halos and peaks at $\sim10^{12} M_{\odot}$. 

\begin{figure*}\centering
\includegraphics[height=2.9in,width=3.45in]{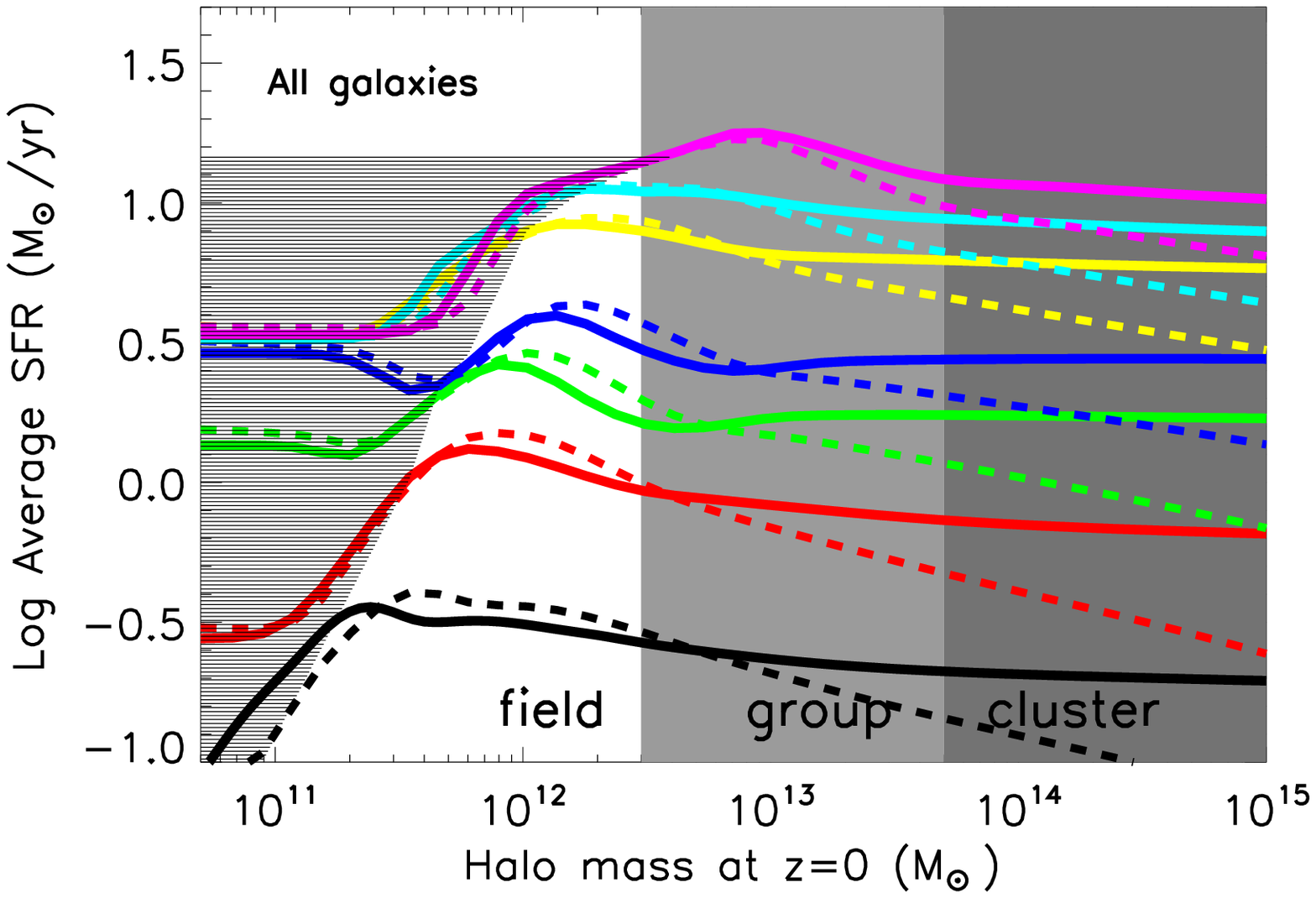}
\includegraphics[height=2.9in,width=3.45in]{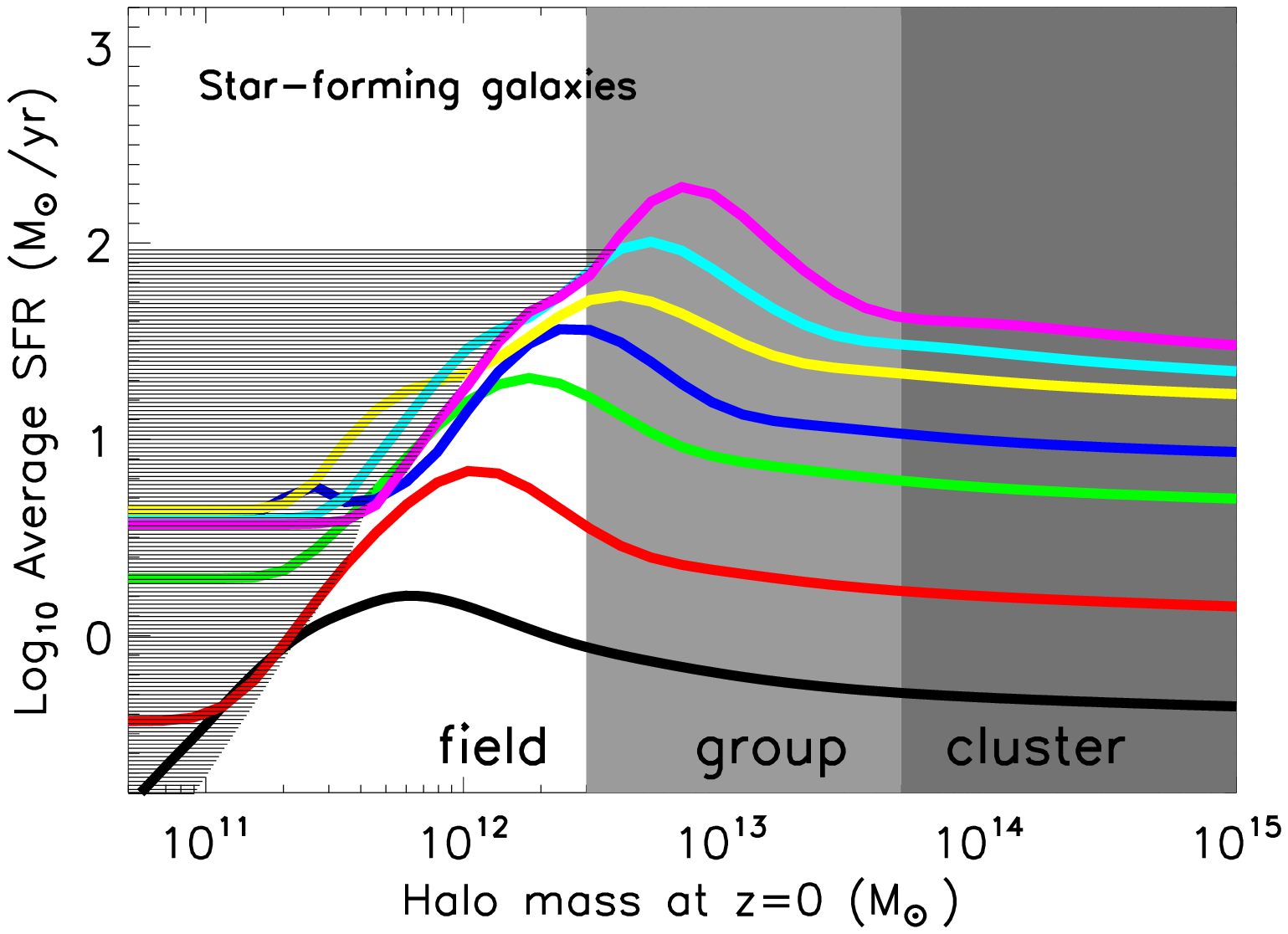}
\caption{Left: The average SFR as a function of halo mass at the present day. $M_{\rm h}$ is evolved to $z=0$ using the halo accretion history from numerical simulations (Fakhouri et al. 2010). The solid/dashed lines correspond to the $\psi$ - $M_{\rm h}$ relation derived from {\bf Scenario A} / {\bf B}.  Along any vertical line, we can trace the evolution of the SFR in the same halo. The dark grey / light grey / white region indicates $M_{\rm h}$  range typically associated with cluster / group /  field environment. For clarity, error bars are not shown here. Right: Similar to the left panel but for star-forming galaxies only. It is clear that the most actively star-forming galaxies at $z\sim2$ reside in group-like environment and they evolve into quiescent galaxies in groups at the present-day.} 
\label{fig:SFR_HM_z0}
\end{figure*}

In Fig.~\ref{fig:SFR_HM_z0}, $M_{\rm h}$ is evolved to $z=0$ using the halo mass accretion history derived from numerical simulations (Fakhouri et al. 2010),  i.e. Eq. (6). So we can trace the star-formation history in the same halo along any vertical line. One can read off the evolutionary sequence of different populations of galaxies. Galaxies that are forming stars most actively at $z\sim2$ have evolved into populations that reside in group-like environments at the present day and galaxies that are forming stars mostly actively in the present-day generally reside in field environment. This explains the reversal of the SFR - density relation at high redshift first presented in Elbaz et al. (2007) and strongly supports previous claims that the most powerful starbursts at $z\sim2$ (i.e. sub-mm galaxies) have evolved into today's elliptical galaxies in dense environment (e.g. Lilly et al. 1999; Smail et al. 2004; Swinbank et al. 2006). It is worth pointing out that our results on the redshift evolution of the average SFR as a function of halo mass are in good qualitative agreement with some recent results in the literature (Behroozi et al. 2012; Moster et al. 2013). A detailed quantitative comparison (e.g., the impact of different methodology, different observations used to constrain the empirical model etc.) is beyond the scope of this paper.

\section{Discussions and conclusions}

In the last ten years there has been an explosion of spectroscopic and multi-wavelength photometric data charting the star-formation history and stellar mass build-up over a large fraction of cosmic time. And now the advent of Herschel allows us to reliably probe the obscured star-formation activity in large numbers of high-$z$ galaxies. In the near future, powerful space- and ground-based facilities will dramatically increase sample size and allow robust measurements of galaxy properties to be made at even higher redshift.

%Error bars include both the variation in the SFR distribution as a function of stellar mass and the uncertainty in the stellar-to-halo mass relation. 

In this paper, we present an extended halo model (EHM) of galaxy evolution which links stellar mass ($m_*$) and SFR of galaxies to their underlying host halo mass ($M_{\rm h}$) from the local Universe to $z\sim2$. While the empirical relation between $m_*$ and $M_{\rm h}$ has been constructed based on observations before, this is the first time the relation between $\psi$ and $M_{\rm h}$ has been constructed from observational data over $80\%$ of cosmic time. The {\it Herschel}-SPIRE observations obtained as part of the {\it Herschel} Multi-tiered Extragalactic Survey (HerMES) is crucial for obtaining accurate SFR estimates for dusty star-forming galaxies at high $z$. 

The EHM is built through two steps:

\begin{itemize}

\item 
First, we build the CSMF $\Phi(m_*|M_{\rm h})$, which specifies the average number of galaxies as a function of $m_*$ in a halo of a given mass. The CSMF, by construction, fits the SMF and the projected correlation functions as a function of stellar mass in the local Universe and the SMFs in various redshift slices in the distant Universe. The predicted clustering properties from our best-fit CSMF as a function of redshift also agree reasonably well with the measured correlation functions at high $z$ (modulo integral constraint and cosmic variance effect). 

\item 
Second, we extend the CSMF to  the joint distribution in $\psi$ and $m_*$ as a function of halo mass, $\Phi(\psi, m_*|M_{\rm h})$, by incorporating the distribution of SFR at fixed $m_*$. We have used two scenarios in building $\Phi(\psi, m_*|M_{\rm h})$. {\bf Scenario A} assumes that $m_*$ plays the most important role in determining the SFR distribution of galaxies and the effect of $M_{\rm h}$ at fixed $m_*$ is negligible. {\bf Scenario B} assumes that the SFR distribution at fixed $m_*$ has a power-law dependence on $M_{\rm h}$. The difference in the resulting $\psi$ - $M_{\rm h}$ relation is small between the two different scenarios and does not affect the main conclusions presented in the paper.

\end{itemize}

Combining the halo accretion history from numerical simulations and the 2-D distribution of $m_*$ and $\psi$ as a function of $M_{\rm h}$ in various redshift slices $\Phi(\psi, m_*|M_{\rm h}, z)$, we can trace the stellar mass growth and the evolution of SFR in different halos.  Our most important findings are: 
\begin{itemize}

\item
The intensity of the star-forming activity in halos in the probed mass range has steadily decreased over time, dropping by over one order of magnitude from $z\sim2$ to $z\sim0$; 

\item 
At each redshift, halos in the mass range between a few times $10^{11}M_{\odot}$ and a few times $10^{12} M_{\odot}$ are the most efficient at hosting star formation, consistent with the optimum halo mass scale for star formation predicted from numerical simulations;  

\item 
The peak of SFR as well as the peak of the stellar-to-halo mass ratio (a measure of the integrated star-formation efficiency) shifts to lower mass halos as redshift decreases; 

\item 
Galaxies that are forming stars most actively at $z\sim2$ have evolved into quiescent galaxies in group-like environments at the present day. 

\end{itemize}
 
To further constrain the physical processes responsible for the $\psi$ - $M_{\rm h}$ relation and its evolution with redshift, future work is needed to investigate the role of three main suspects: molecular gas content and evolution, feedback from central massive black holes and environmental effects on star formation. 
 The advent of the Atacama Large Millimeter/submillimeter Array (ALMA), the Expanded Very Large Array (EVLA; Perley et al. 2011) and the Northern Extended Millimeter Array (NOEMA) means that we are now in a position to be able to measure the evolution of the molecular gas content in a statistically significant sample of galaxies with moderate SFRs. The feedback from growing black holes may also impact the star-formation activity in massive halos ($M_{\rm h}>10^{13} M_{\odot}$), as required in order to reproduce the observed stellar mass and luminosity functions of galaxies in numerical simulations and semi-analytic models (e.g., Bower et al. 2006, 2008; Croton et al. 2006). We will extend the EHM to include the empirical relation between AGNs and halo mass in a future paper to statistically investigate star formation - black hole co-evolution. Finally, the impact of environment can be studied through galaxy group and cluster catalogues over a large redshift range and will be presented in a separate paper.

\section*{ACKNOWLEDGEMENTS}
We thank the anonymous referee for constructive comments. We also thank Cheng Li for providing the SDSS clustering measurements and the stellar mass function, Rachel Mandelbaum for the stellar mass $m_*$ - halo mass $M_{\rm h}$ relation from weak lensing,  Surhud More for the $m_*$ - $M_{\rm h}$ relation from satellite kinematics, Alexi Leauthaud for the $m_*$ - $M_{\rm h}$ relation, Samir Salim for the conditional dependence of SSFR on $m_*$ in the local Universe, Benjamin Moster for useful discussions on the CSMF, Anthony Lewis for useful discussion on MCMC methods and Peder Norberg for general discussions.

LW acknowledges support from UK's Science and Technology Facilities Council grant ST/F002858/1 and an ERC StG grant (DEGAS-259586). SJO is supported by UK's Science and Technology Facilities Council grant ST/F002858/1. The data presented in this paper will be released through the Herschel database in Marseille HeDaM (hedam.oamp.fr/herMES). SPIRE has been developed by a consortium of institutes led by Cardiff Univ. (UK) and including Univ. Lethbridge (Canada); NAOC (China); CEA, LAM (France); IFSI, Univ. Padua (Italy); IAC (Spain); Stockholm Observatory (Sweden); Imperial College London, RAL, UCL-MSSL, UKATC, Univ. Sussex (UK); Caltech, JPL, NHSC, Univ. Colorado (USA). This development has been supported by national funding agencies: CSA (Canada); NAOC (China); CEA, CNES, CNRS (France); ASI (Italy); MCINN (Spain); SNSB (Sweden); STFC (UK); and NASA (USA).

\appendix
\section{The conditional stellar mass function}
\label{first appendix}

Motivated by studies of galaxy groups (Yang et al. 2005), we can divide the CSMF into that of central and satellite galaxies,
\begin{equation}
\Phi(m_{\star}|M_{\rm h}) = \Phi_{\rm cen}(m_{\star}|M_{\rm h}) + \Phi_{\rm sat}(m_{\star}|M_{\rm h}),
\end{equation}
where $\Phi_{\rm cen}(m_*|M_{\rm h})$ and $\Phi_{\rm sat}(m_*|M_{\rm h})$ specify the number of central and satellite galaxies as a function of $m_*$ at fixed $M_{\rm h}$ respectively. A log-normal distribution is used to model the CSMF of central galaxies, 
\begin{equation}
\Phi_{\rm cen}(m_{\star}|M_{\rm h}) = \frac{1}{\sqrt{2\pi}\ln 10 m_{\star} \sigma_{\rm c}} \exp{\left[-\frac{\log^2(m_{\star}/m_{\rm c})}{2\sigma_c^2}\right]},
\end{equation}
where $m_{\rm c}(M_{\rm h})$ is the mean stellar mass of a central galaxy in a halo of mass $M_{\rm h}$ and $\sigma_{\rm c}=0.2$ dex is the standard deviation. Following Moster et al. (2010), 
$m_{\rm c}(M_{\rm h})$ is parametrized as, 
\begin{equation}
m_{\rm c}(M_{\rm h}) = 2M_{\rm h} \left(\frac{m_{\rm c}}{M}\right)_0 \left[  \left( \frac{M_{\rm h}}{M_{1c}}\right)^{-\beta_{\rm c}} + \left( \frac{M_{\rm h}}{M_{1c}} \right)^{\gamma_{\rm c}} \right]^{-1},
\end{equation}
where $ \left(\frac{m_{\rm c}}{M}\right)_0$ is the overall normalisation, $\beta_{\rm c}$ and $\gamma_{\rm c}$ controls $m_{\rm c}(M_{\rm h})$ at the low and high halo mass end respectively, and $M_{1c}$ is the characteristic halo mass scale. A modified Schechter function is used to model the CSMF of satellites,
\begin{equation}
\Phi_{\rm sat}(m_{\star}|M_{\rm h}) = \frac{\Phi_s^*}{m_s}\left(\frac{m_{\star}}{m_s}\right)^{\alpha_s} \exp{\left[-\left(\frac{m_{\star}}{m_s}\right)^2\right]}, 
\end{equation}
where $\alpha$ is the low-mass end slope, 
\begin{equation}
\alpha = \alpha_0 + \alpha_s \times \log  \left( \frac{M_{\rm h}}{M_{\odot}}\right)
\end{equation}
$\Phi_s^*$ is the normalisation,
 \begin{equation}
\Phi_s^*(M_{\rm h}) = \Phi_0 \left( \frac{M_{\rm h}}{M_{\odot}}\right),
 \end{equation}
 and $m_s$ is the characteristic stellar mass in the distribution of satellites,  
\begin{equation}
m_{\rm s}(M_{\rm h}) = 2M_{\rm h} \left(\frac{m_{\rm s}}{M}\right)_0 \left[  \left( \frac{M_{\rm h}}{M_{1s}}\right)^{-\beta_{\rm s}} + \left( \frac{M_{\rm h}}{M_{1s}} \right)^{\gamma_{\rm s}} \right]^{-1}
\end{equation}
which has the same functional form as $m_{\rm c}(M_{\rm h})$.
 
Equipped with the CSMF, we can calculate the abundance and clustering of galaxies. For example, the SMF can be derived as follows,
\begin{equation}
\Phi(m_*) = \int_0^{\infty} \Phi(m_*|M_{\rm h}) n(M_{\rm h}) dM_{\rm h}, 
\end{equation}
where $n(M_{\rm h})$ is the halo mass function (HMF). In this paper, we use the HMF from Tinker et al. (2008). The galaxy power spectrum as a function of $m_*$ is
\begin{equation}
P_{\rm gal}(k|m_*) = P_{\rm 1h}(k|m_*) + P_{\rm 2h}(k|m_*).
\end{equation}
 The 1-halo term comes from galaxy pairs in the same halo,
\begin{eqnarray}
P_{\rm 1h}(k|m_{\star}) & = & \frac{1}{\Phi(m_*)^2} \int n(M_{\rm h}) [\Phi_{\rm sat}(m_{\star}|M_{\rm h})^2 u_{\rm g}(k|M_{\rm h})^2 + \nonumber\\
               &&2 \Phi_{\rm cen}(m_{\star}|M_{\rm h }) \Phi_{\rm sat}(m|M_{\rm h}) u_{\rm g}(k|M_{\rm h})] dM_{\rm h}.
\end{eqnarray} 
Here $u_{\rm g}(k|M_{\rm h})$ is the normalised Fourier transform of the galaxy density distribution within a halo of mass $M_{\rm h}$,
assumed to be an NFW profile (Navarro, Frenk \& White 1997) truncated at the virial radius.
The 2-halo term comes from galaxy pairs in separate halos,
\begin{eqnarray}
&P_{\rm 2h}(k|m_{\star})&=  \left[\int dM_{\rm h} n(M_{\rm h}) b(M_{\rm h}) \frac{\Phi(m_{\star}|M_{\rm h})}{\bar{n}(m_{\star})} u_{\rm g}(k|M_{\rm h})\right]^2 \nonumber \\
 &&P^{\rm lin}(k)
\end{eqnarray}
Here $P^{\rm lin}(k)$ is the linear dark matter power spectrum, $b(M_{\rm h})$ is the bias factor as a function of $M_{\rm h}$ (Tinker et al. 2008). The projected correlation function at a given stellar mass is
\begin{equation}
w_p(r|m_{\star}) = \int \frac{kdk}{2\pi} P_{\rm gal}(k|m_{\star}) J_0(kr),
\end{equation}
where $J_0(x) = \sin(x)/x$ is the zeroth-order Bessel function.

There are a total of 11 parameters in the CSMF of the local Universe. 
We make use of Markov Chain Monte Carlo (MCMC) methods to derive the posterior PDF for all parameters by fitting to the observed abundance and clustering properties. Specifically, we use MCMC to minimise the reduced chi-squared 
\begin{eqnarray}
\chi_{\rm r}^2 & = &\frac{1}{N_{\Phi}}\Sigma_1^{N_{\Phi}} [(\Phi_{\rm CSMF}(m_{\star}) - \Phi_{\rm obs}(m_{\star}))/\sigma_{\Phi}]^2   + \nonumber\\
&&\frac{1}{N_{\rm s}}  \Sigma_1^{N_{\rm s}} \frac{1}{N_{\rm r}} \Sigma_1^{N_{\rm r}} [(w_{p_{\rm CSMF}} - w_{p_{\rm obs}})/\sigma_{w_p}]^2,
\end{eqnarray}
where $N_{\Phi}$ is the number of data points in the SMF, $N_{\rm r}$ is the number of data points in each projected correlation function and $N_{\rm s}$ is the total number of correlation functions. The best-fit value and the standard deviation for each parameter is listed in Table A1. The correlation matrix of the parameters in the CSMF of the local Universe is shown in Table A2.

\begin{table}
\caption{Parameters in the CSMF of the local Universe. The first four parameters describe the distribution of central galaxies as a function of $m_*$ at fixed $M_{\rm h}$, which is assumed to follow a log-normal distribution. The last seven parameters describe the distribution of satellites as a function of $m_*$ at fixed $M_{\rm h}$, which is  assumed to follow a modified Schechter function.}\label{table:CMF}
\begin{tabular}[pos]{llll}
\hline
parameter                   & best-fit             & error   & description\\
\hline
$\log M_{1c}$            & 11.70              & 0.49   & characteristic halo mass\\
                                                                & &   &  in the $m_*/M_{\rm h}$ ratio\\
 $(m_c/M)_0$            &  1.73           &  0.07   & overall normalisation\\
$\beta_c$                   &  1.16                 &  0.06   & power-law slope of $m_*/M_{\rm h}$ \\
 && &   at the low-mass end\\
$\gamma_c$             &  0.71               &  0.03   & power-law slope of $m_*/M_{\rm h}$ \\
&&&  at the high-mass end\\
\hline
 $\log M_{1s}$          &  12.62                 & 0.55     & characteristic halo mass \\
 &&   &  in the $m_*/M_{\rm h}$ ratio\\
 $(m_s/M)_0$           &  2.32       &   0.13     & overall normalisation\\
$\beta_s$                  &  2.38               & 0.33          &power-law slope of $m_*/M_{\rm h}$\\
 && &   at the low-mass end\\
$\gamma_s$            &  0.97              & 0.05         & power-law slope of $m_*/M_{\rm h}$\\
&&&  at the high-mass end\\
$-\log \Phi_0$          &  13.11               & 0.54  &overall normalisation in\\
&&&the number of satellites\\
$-\alpha_0$              &  0.28                & 0.11        &  power-law slope in $N_{\rm sat}$   \\
$-\alpha_s$              &  0.06                & 0.01       &  power-law slope in $N_{\rm sat}$   \\
&&& at the low-mass end\\
\hline
\end{tabular}
\end{table}

\begin{table*}
\caption{The correlation matrix of the parameters describing the CSMF of the local Universe.}\label{table:CMF}
\begin{tabular}[pos]{llllllllllll}
\hline
        & $\log M_{1c}$   &  $(m_c/M)_0$ & $\beta_c$ & $\gamma_c$   & $\log M_{1s}$ &  $(m_s/M)_0$ & $\beta_s$  & $\gamma_s$ & $-\log \Phi_0$ & $\alpha_0$&  $\alpha_s$  \\
\hline
$\log M_{1c}$            &         1.00 & -0.66   & -0.70  &   0.79 & -0.54  &  0.56 &   -0.04 &   -0.44 &   -0.74 &   -0.15 &     0.11\\ 
 $(m_c/M)_0$            &           &  1.00    &  0.44  &  -0.11 &  0.43  & -0.48 &    0.06 &    0.17 &    0.59 &    0.19 &    -0.16\\ 
$\beta_c$                   &           &         &    1.00 &  -0.55 &  0.13  & -0.15 &   -0.12 &    0.11 &    0.48 &   -0.11 &     0.15\\ 
$\gamma_c$             &            &         &        &    1.00 & -0.38  &  0.37 &   -0.01 &   -0.43 &   -0.50 &   -0.06 &    0.03\\ 
$\log M_{1s}$               &           &         &        &        &    1.00 & -0.99 &   -0.02 &    0.81 &    0.62 &    0.11 &   -0.07\\ 
 $(m_s/M)_0$          &           &         &        &        &        &   1.00 &   -0.04 &   -0.72 &   -0.62 &   -0.13 &    0.09\\ 
 $\beta_s$           &           &         &        &        &        &       &    1.00  &  -0.18  &   -0.01 &    0.21 &    -0.21 \\ 
 $\gamma_s$                  &           &         &        &        &        &       &         &     1.00 &    0.58 &    0.05 &   -0.01\\ 
 $-\log \Phi_0$            &          &         &        &        &        &       &         &         &    1.00  &    0.17 &   -0.09\\ 
 $\alpha_0$          &          &         &        &        &        &       &         &         &         &     1.00 &    -0.99\\ 
 $\alpha_s$             &          &         &        &        &        &       &         &         &         &         &     1.00\\ 
\hline 
\end{tabular}
\end{table*}

To add in the redshift evolution of the CSMF, we adopt the following parametrisation to describe the evolving $m_*$ - $M_{\rm h}$ relation, following Moster et al. (2010). The evolution in the characteristic halo mass scale is parameterised as, 
\begin{equation}
\log M_1(z) = (1+z)^{\mu}  \times \log M_1|_{z=0}. 
\end{equation}
And the overall normalisation in the stellar-to-halo mass ratio is parameterised as,
\begin{equation}
\left(  \frac{m}{M}  \right)_0 (z) = (1+z)^{\nu} \times \left(  \frac{m}{M}  \right)_0 |_{z=0}
\end{equation}
Finally, the power-law slope at the high-mass and low-mass end are parameterised as
\begin{equation}
\gamma(z) = (1+z)^{\gamma_1} \times \gamma|_{z=0}, 
\end{equation}
and
\begin{equation}
\beta(z) = \beta|_{z=0} + \beta_1 \times z,
\end{equation}
respectively. We use the SMF in the high-$z$ Universe to constrain the redshift evolution of the $m_*$ - $M_{\rm h}$ relation. The best-fit value and the standard deviation for each parameter is listed in Table A3. The correlation matrix of the 4 parameters used to describe the redshift evolution of the CSMF is shown in Table A4. We have also tried to use 8 evolution parameters to allow different redshift evolution for the central and satellite population. However, the parameters are highly correlated and the uncertainties on these parameters are very large from MCMC chains.

\begin{table}
\caption{The redshift evolution parameters in the CSMF.}\label{table:CMF}
\begin{tabular}[pos]{lll}
\hline
parameter                   & best-fit             & error   \\
\hline
$\mu$                          &  0.028      & 0.010                    \\
$\nu$                          &   0.780   &     0.176                      \\
$\beta_1$                   &  0.079     &           0.133   \\
$\gamma_1$             &  -0.061  &        0.268   \\
\hline
\end{tabular}
\end{table}

\begin{table}
\caption{The correlation matrix of the redshift evolution parameters in the CSMF.}\label{table:CMF}
\begin{tabular}[pos]{lllll}
\hline
                   & $\mu$               & $\nu$     &  $\beta_1$ & $\gamma_1$\\
\hline
$\mu$                          &      1.00    & 0.57  &   -0.64  &   0.76\\
$\nu$                          &                  & 1.00  &   -0.02  &    0.85\\
$\beta_1$                   &                 &           &     1.00 &    -0.27\\
$\gamma_1$             &                 &           &              &     1.00\\
\hline
\end{tabular}
\end{table}

\section{The projected two-point correlation function}

%The spatial two-point correlation function is often used to study galaxy clustering. It is defined as the probability of finding a galaxy pair at a given separation, in excess of that in a random Poisson distribution. We use the Landy \& Szalay (1993) estimator
%\begin{equation}
%\xi (r_p, \pi) = \frac{1}{RR}\left[DD\left(\frac{n_R}{n_D}\right)^2-2DR\left(\frac{n_R}{n_D}\right)+RR\right].
%\end{equation}
%Here $r_p$ and $\pi$ are the separations perpendicular and parallel to the line of sight, $n_D$ and $n_R$ are the mean densities of the galaxy and random catalogues respectively. $DD (r)$, $DR (r)$ and $RR (r)$ are numbers of weighted galaxy-galaxy pairs, galaxy-random pairs and random-random pairs at separation $r$ respectively. For volume-limited samples, the weight applied to each galaxy is 1.  When generating random catalogues for clustering calculation, the angular distribution of random galaxies is modulated by an angular mask, which is generated using the optical flags to take into account the selection effect. The projected correlation function which integrates out the effects of peculiar velocities and uncertainties in the photometry redshifts can be derived by integrating $\xi(r_p, \pi)$ along $\pi$,
%\begin{equation}
%w_p(r_p) = 2 \int_0^{\infty} \xi(r_p, \pi)d\pi.
%\end{equation}

\begin{figure}\centering
\includegraphics[height=2.4in,width=2.4in]{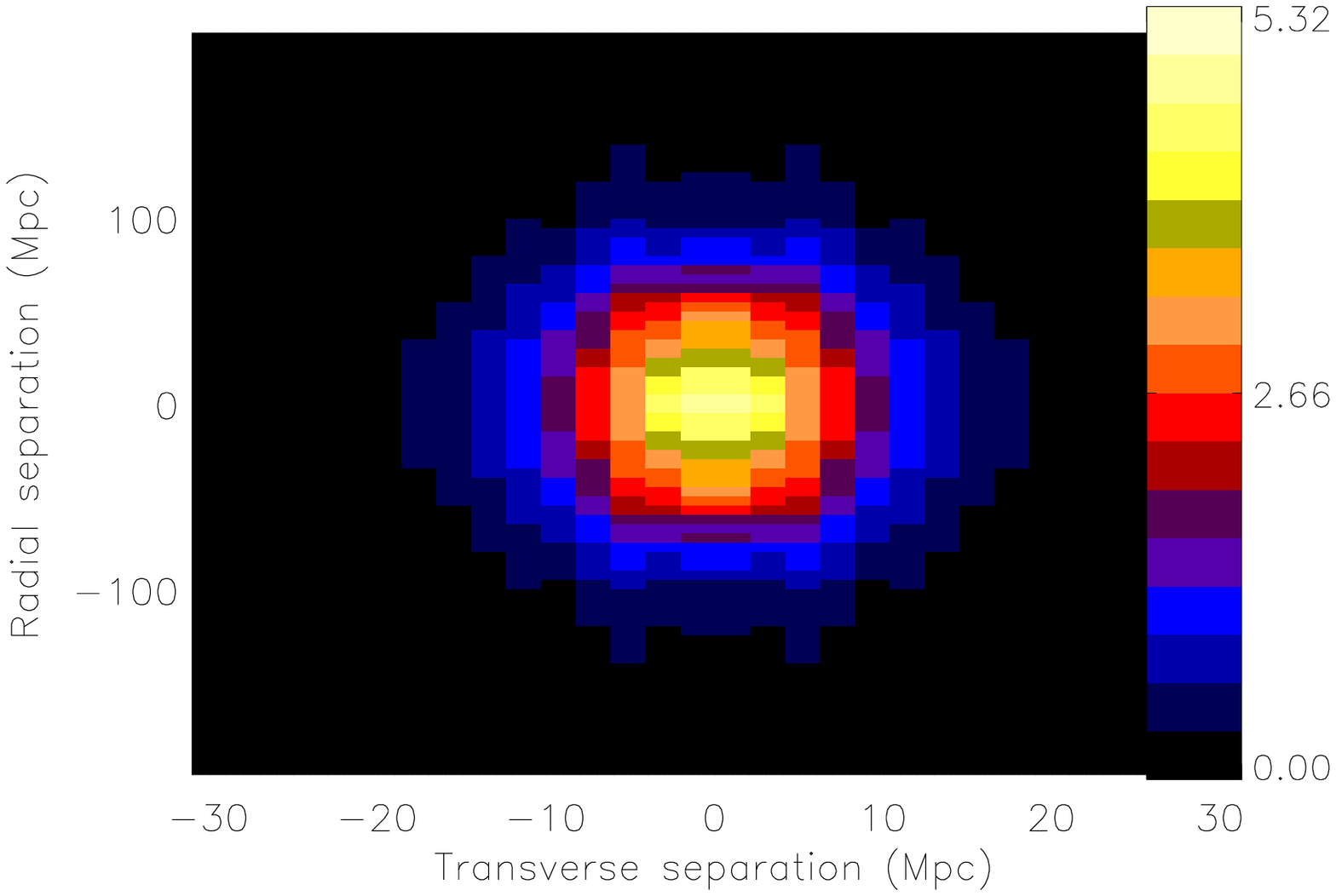}
\caption{The redshift-space correlation function $\xi(r_p, \pi)$. The data from the first quadrant are repeated with reflection in both axes.The signal along the radial direction has been smoothed by a box filter of length 20 Mpc.}
\label{fig:xi_LS}
\end{figure}

The spatial two-point correlation function is often used to study galaxy clustering. It is defined as the probability of finding a galaxy pair at a given separation, in excess of that in a random Poisson distribution. We use the Landy \& Szalay (1993) estimator
\begin{equation}
\xi (r_p, \pi) = \frac{1}{RR}\left[DD\left(\frac{n_R}{n_D}\right)^2-2DR\left(\frac{n_R}{n_D}\right)+RR\right].
\end{equation}
Here $r_p$ and $\pi$ are the separations perpendicular and parallel to the line of sight, $n_D$ and $n_R$ are the mean densities of the galaxy and random catalogues respectively. $DD (r)$, $DR (r)$ and $RR (r)$ are numbers of weighted galaxy-galaxy pairs, galaxy-random pairs and random-random pairs at separation $r$ respectively. For volume-limited samples, the weight applied to each galaxy is 1.  When generating random catalogues for clustering calculation, the angular distribution of random galaxies is modulated by an angular mask, which is generated using the optical flags to take into account the selection effect. In Fig.~\ref{fig:xi_LS}, we plot the $\xi(r_p, \pi)$ of galaxies with $m_*>10^{9.8}M_{\odot}$  in the redshift bin $z1=[0.2, 0.5]$, averaged over COSMOS and EGS. The signal from the first quadrant is repeated with reflection in both axes. In the absence of peculiar velocity and redshift error, $\xi(r_p, \pi)$ should be isotropic. The elongation of the signal along $\pi$ leads to a reduction in the clustering amplitude. The problem can be overcome by integrating $\xi(r_p, \pi)$ along $\pi$ to derive the projected correlation function, 
\begin{equation}
w_p(r_p) = 2 \int_0^{\infty} \xi(r_p, \pi)d\pi.
\end{equation}
Fig.~\ref{fig:xi_LS} also indicates that integrating $\xi(r_p, \pi)$ out to $\pi=160$ Mpc should capture all correlated signal.

% The x-axis is the transverse separation (Mpc) and the y-axis is the radial separation (Mpc). 

\end{document}